\documentclass{aa}
\usepackage{epsfig,rotate}

\def\mum {$\mu$m}

\begin{document}

\thesaurus{08(02.19.1;              
              09.01.1;              
              09.09.1 Orion Peak 1; 
              09.13.2;              
              13.09.4)              
}
\title{
ISO-SWS Observations of OMC-1: H$_2$ and Fine Structure Lines\thanks{
Based on observations with ISO, an ESA project with instruments funded
by ESA Member States (especially the PI countries: France, Germany, The
Netherlands, and the United Kingdom) and with the participation of ISAS
and NASA.}
}

\author{
Dirk Rosenthal\inst{1}    \and
Frank Bertoldi\inst{2}    \and
Siegfried Drapatz\inst{1}
}

\institute{
Max-Planck-Institut f\"{u}r extraterrestrische Physik,
Giessenbachstrasse, D-85740 Garching, Germany,\\ Rosenthal@MPE.MPG.de \and
Max-Planck-Institut f\"ur Radioastronomie,
Auf dem H\"ugel 69, D-53121 Bonn, Germany, Bertoldi@MPIfR-Bonn.MPG.de
}

\offprints{D. Rosenthal}
\date{Received ; accepted }
\authorrunning{D. Rosenthal et al.}
\maketitle


\begin{abstract}
  
  Using the Short-Wavelength-Spectrometer on the Infrared Space
  Observatory (ISO), we obtained near- and mid-infrared spectra toward
  the brightest H$_2$ emission peak of the Orion OMC-1 outflow. A wealth
  of emission and absorption features were detected, dominated by 56
  $\rm H_2$ ro-vibrational and pure rotational lines reaching from H$_2$
  0--0 S(1) to 0-0~S(25).  The spectra also show a number of H~{\sc i}
  recombination lines, atomic and ionic fine structure lines, and
  molecular lines of CO and $\rm H_2O$.  Between 6 and 12 $\rm \mu m$
  the emission is dominated by PAH features. 
  
  The extinction toward the molecular and atomic line emitting regions
  is estimated from relative line intensities, and it is found that the
  H$_2$ emission arises from within the OMC-1 cloud at an average K-band
  extinction of 1.0 mag, whereas the atomic hydrogen emission and much
  of the fine structure emission comes from the foreground H~{\sc ii}
  region and its bounding photodissociation front.
  
  The total H$_2$ luminosity in the ISO-SWS aperture is estimated at
  $(17 \pm 5)~\rm L_{\sun}$, and extrapolated to the entire outflow,
  $(120 \pm 60)~\rm L_{\sun}$.  The H$_2$ level column density
  distribution shows no signs of fluorescent excitation or a deviation
  from an ortho-to-para ratio of three.  It shows an excitation
  temperature which increases from about 600~K for the lowest rotational
  and vibrational levels to about 3200~K at level energies $E(v,J)/k >
  14\,000$~K. No single steady state shock model can reproduce the
  observed H$_2$ excitation.  The higher energy H$_2$ levels may be
  excited either thermally in non-dissociative J-shocks, through
  non-thermal collisions between fast ions and molecules with H$_2$ in
  C-shocks, or they are pumped by newly formed H$_2$ molecules.  The
  highest rotational levels may be populated by yet another mechanism,
  such as the gas phase formation of H$_2$ through H$^-$.


\end{abstract}


\section{Introduction}
 
The Orion molecular cloud, OMC-1, located behind the Orion M42 Nebula at
a distance of $\sim$450~pc (Genzel \& Stutzki \cite{gen89}), is the
best-studied massive star forming region.  This cloud embeds a
spectacular outflow arising from some embedded young stellar object,
which can possibly be identified as the radio source ``I'' 0.49 arcsec
south of the infrared source IRc2-A (Menten \& Reid \cite{men95};
Dougados et al. \cite{dou93}).  The outflow shocks the surrounding
molecular gas, thereby giving rise to the strongest H$_2$ infrared line
emission appearing in the sky (Fig.~\ref{schultz}).  Peak~1 (Beckwith et
al. \cite{bec78}) is the brighter of the two H$_2$ emission lobes of the
outflow.  Although the outflow has been studied extensively for nearly
two decades, the nature of the excitation mechanism remains unclear.

Molecular hydrogen, through its infrared rotational and
rotation-vibrational transitions, is an important coolant in shocks and
photodissociation regions, and thereby a particularly well suited tracer
of the flourescently- and/or shock-excited gas.  The Short Wavelength
Spectrometer (SWS, de Graauw et al. \cite{deg96}) aboard the Infrared
Space Observatory (ISO, Kessler et al. \cite{kes96}) offered the first
opportunity to observe pure rotational and rotation-vibrational H$_2$
lines from 2.4 $\mu$m to 28 $\mu$m with one instrument, unhindered by
the Earth's atmosphere.

In this paper, we present a comprehensive set of intensities for 56
H$_2$ near- and mid-infrared lines we observed with the ISO-SWS.  These
observations trace populations of energy levels ranging from $E/k=$1015~K
to 43\,000~K.  The redundancy of the H$_2$ level determinations provides
information on the average gas excitation along the line of sight over
an unprecedented range.  This sheds new light on the possible excitation
mechanisms in the IRc2 outflow.

We here concentrate on the interpretation of the $\rm H_2$ and the
atomic and ionic fine structure line emission, whereas a detailed
discussion of the CO and $\rm H_2O$ lines will be presented in a
separate paper (Boonman et al.  \cite{boo00}).  In a related paper we
already discussed the detection of HD toward Orion Peak~1 (Bertoldi et
al. \cite{ber99}).

 
\section{Observation}
 
\begin{figure}[htb] 
\begin{center}
\includegraphics[width=1.0\columnwidth]{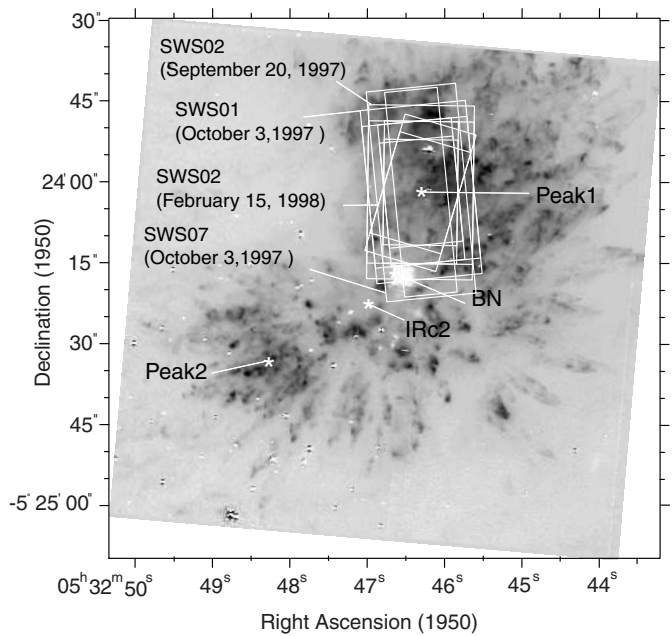}
\caption{
  H$_2$~1-0 S(1) emission of the OMC-1 outflow as seen with the NICMOS
  camera aboard the HST (Schultz et al. 1999).  Overlaid are the various
  apertures of our ISO-SWS observations, which were centered on
  $\alpha_{\rm 2000.} = 5^{\rm h} 35^{\rm m} 13.\! \!^{\rm \, s} 67$, $
  \delta_{\rm 2000.} = -5 \degr 22 \arcmin 8.\! \!  \arcsec 5$, with an
  aperture of $14 \arcsec \times 20 \arcsec $ for $\lambda < 12$ $\rm
  \mu m$, $14 \arcsec \times 27 \arcsec $ at 12 to 27.5\mum, $ 20
  \arcsec \times 27 \arcsec $ at 27.5 to 29\mum, and $20 \arcsec
  \times 33 \arcsec $ at 29 to 45.2\mum.}
\label{schultz}
\end{center}
\end{figure}

We observed OMC-1 in the SWS~01 ($\rm 2.4-45~\mu m$ grating scan) and
SWS~07 (Fabry-P\'{e}rot) modes of the short wavelength spectrometer (de
Graauw et al. \cite{deg96}) on board ISO on October 3, 1997, and in the
SWS~02 ($\approx 0.01 \lambda$ range grating scan) mode on September 20,
1997 and February 15, 1998.  Fig.~\ref{schultz} illustrates the various
aperture orientations with respect to the $\rm H_2$~1-0~S(1) emission at
2.12~$\rm \mu m$ observed with NICMOS on the HST (Schultz et al.
\cite{sch99}).

For the 2.4--45~$\mu$m spectrum of Peak~1 we used the slowest speed,
highest spectral resolution full scan observing mode of the SWS.  Data
reduction was carried out using standard Off Line Processing (OLP)
routines up to the Standard Processed Data (SPD) stage within the SWS
Interactive Analysis (IA) system. Between the SPD and Auto Analysis
Result (AAR) stages, a combination of standard OLP and in-house routines
were used to extract the individual scans as well as for the removal of
fringes. The flux calibration errors range from 5\%{} at 2.4 \mum\ to
30\% at 45\mum\ (SWS Instrument \& Data Manual, Issue 1.0).  The
statistical uncertainties derived from the line's signal to noise ratio
are for most detected lines smaller than the systematic errors due to
flux calibration uncertainties.


\section{Results and Discussion}

\begin{figure*}[htb] 
\begin{center}
\includegraphics[width=2.\columnwidth]{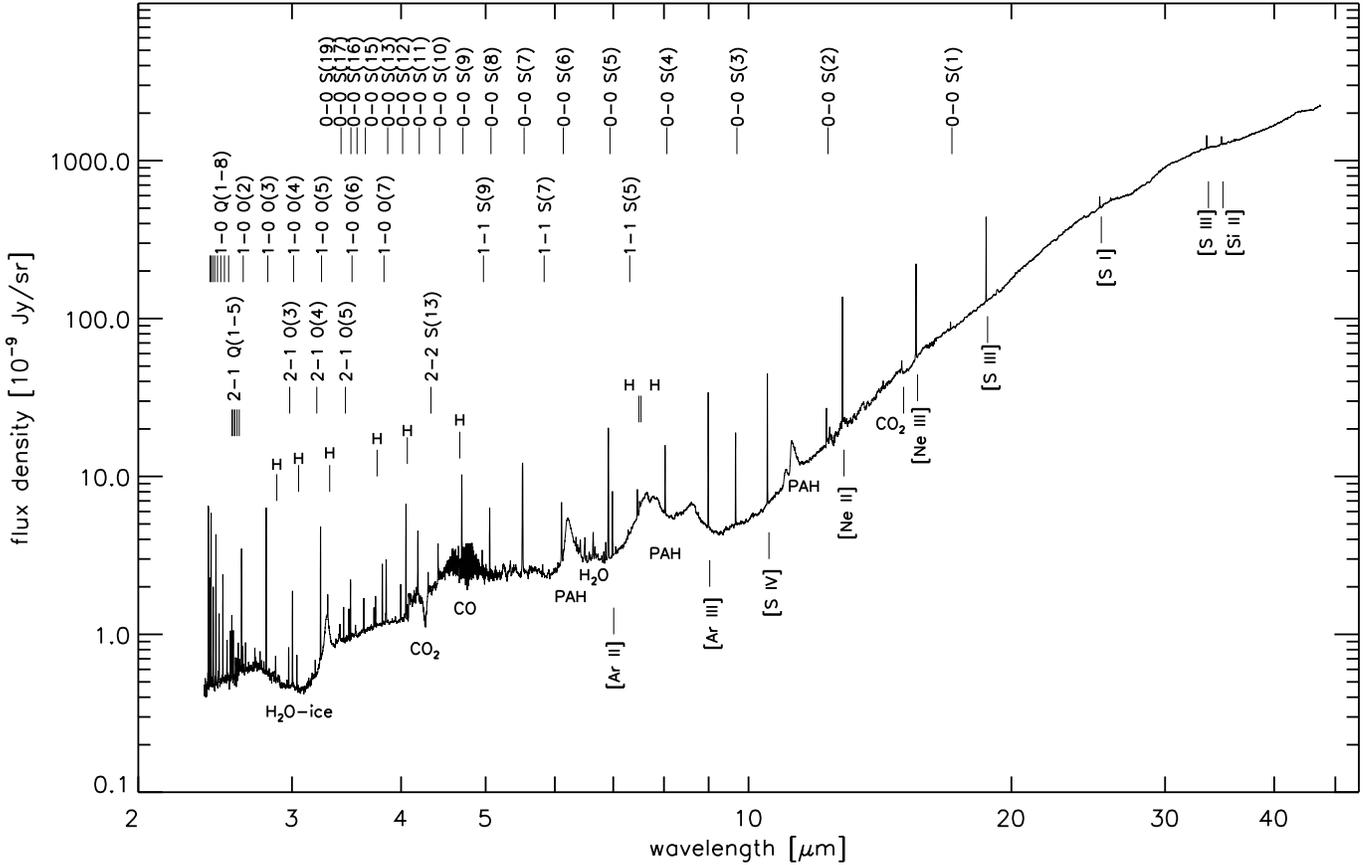} 
\caption{2.4--45 \mum\ spectrum of Orion Peak~1 
  obtained in the SWS~01 grating scan observing mode. Some of
  the detected lines, bands and features are identified. 
  The continuum levels
  of the individual bands, which differ due to aperture changes, were
  adjusted to make the spectrum appear continuous.}
\label{spectrum}
\end{center}
\end{figure*}

Fig.~\ref{spectrum} shows the full SWS~01 spectrum.  Most of the
observed continuum flux is probably coming from the strong Becklin
Neugebauer (BN) source near the edge of the aperture.
Aperture size changes from one band to
another then cause changes in the intercepted continuum which are not
simply proportional to the aperture size. In addition, there is extended
continuum emission all over the outflow.

We normalized the line and continuum fluxes by the aperture size,
assuming that there is uniform surface brightness at least for the line
emission. The exact aperture profiles for the various wavelength bands
is yet to be determined.  Assuming an effective aperture resembling
those shown in Fig.~\ref{schultz} is approximate. The error from this
assumption, and the nonuniform continuum surface brightness cause
additional relative offsets in the continuum fluxes of neighboring bands
of --10\% for the 7--12~$\mu$m band, $-30\%$ for the 12--16~$\mu$m band,
+15\% at 16--19.5~$\mu$m, -5\% at 19.5--27.5~$\mu$m, -7.5\% at 27.5--29.5~$\mu$m,
and +5\% above 29.5~$\mu$m.

\begin{figure*}[h] 
\begin{center}
\includegraphics[width=1.95\columnwidth]{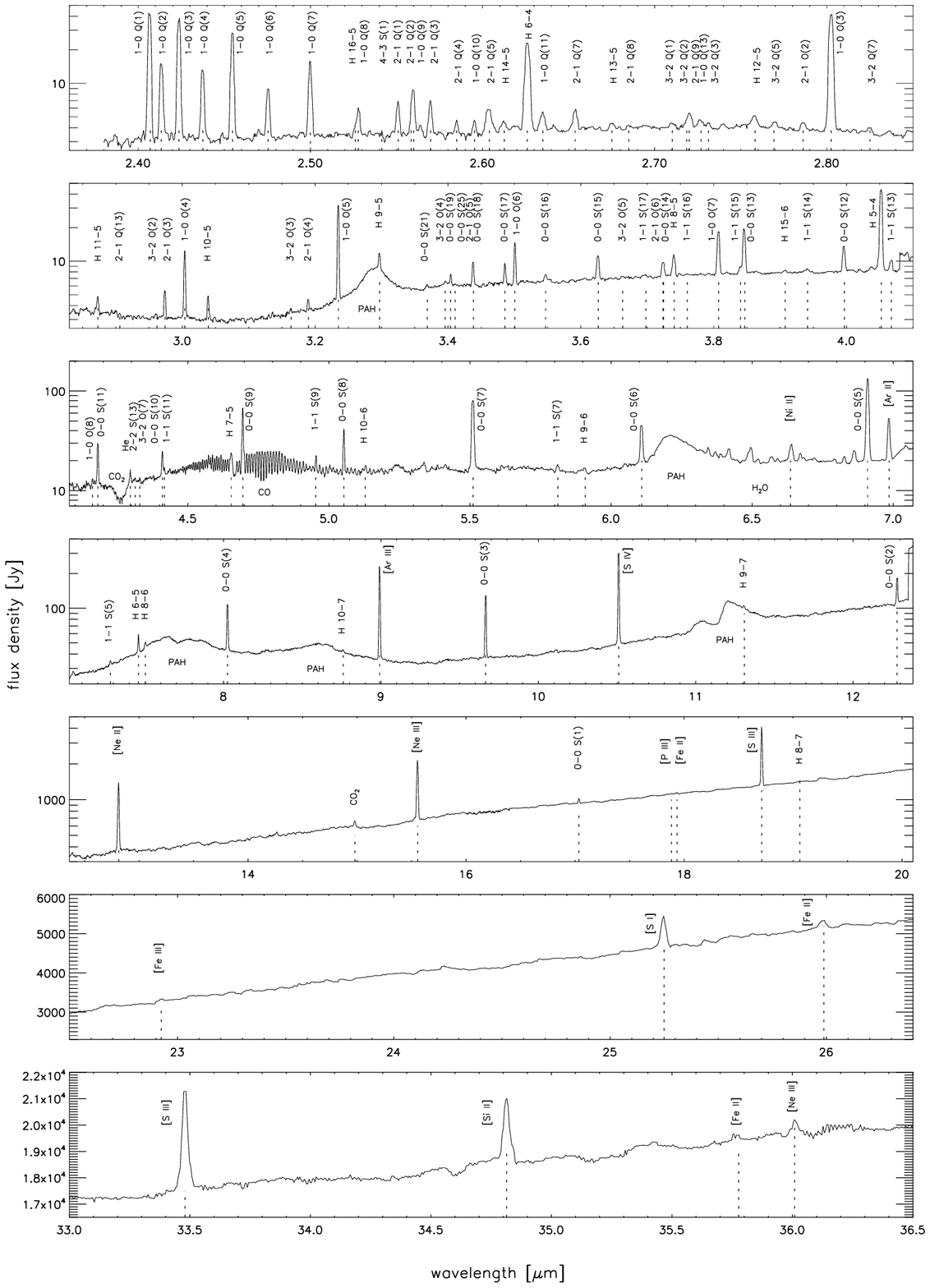} 
\vspace{0.2cm}
\caption{The full scan SWS 01 spectrum of Fig. 2 in detail.}
\label{spec}
\end{center}
\end{figure*}

\begin{figure*}[h] 
\begin{center}
\includegraphics[width=2.0\columnwidth]{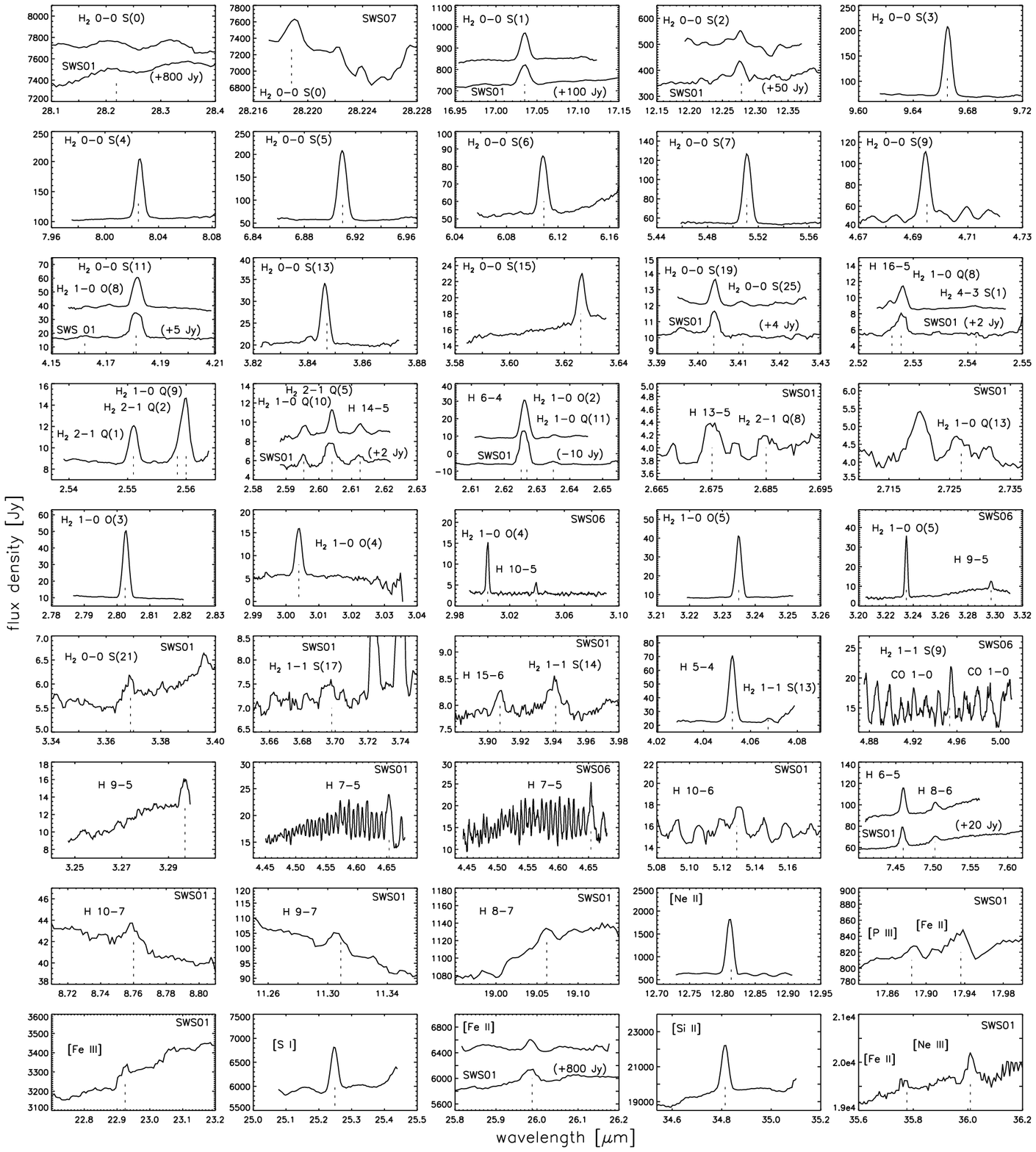}
\vspace{0.5cm}
\caption{Line scans in the SWS 02 or SWS 07/SWS 06 
  modes.  In cases where the flux from the SWS 01 full scan spectrum
  differs much from that found in other observing modes, or where lines
  do not show well in Figs.~\ref{spectrum} and \ref{spec}, we overplot
  the SWS 01 spectrum for comparison.}
\label{multi}
\end{center}
\end{figure*}

Since the observed H$_2$ line intensities result in a smooth
distribution of column densities in the excitation diagram, Fig.
\ref{excit}, the line intensities appear not to be affected much by the
uncertainty of the aperture.  Fig.~\ref{spec} shows the SWS~01
spectrum in more detail.  Fig.~\ref{multi} shows selected lines at
higher spectral resolution from line scan observation in the SWS~02 and
SWS~07/SWS~06 modes.  The SWS~06 grating spectra were simultaneously
recorded with the SWS~07 Fabry-Perot spectra.

The Peak 1 spectrum is dominated by a large number of rotational and
ro-vibrational $\rm H_2$ lines.  The pure rotational lines 
arise from levels with energies
ranging from $E/k=$1015~K for the 0-0~S(1) line to
$E/k=42\,515$~K for the 0-0~S(25) line. They represent gas with
excitation temperatures ranging from 600~K for the low energy levels to
over 3000~K for level energies $E/k > 14\,000$~K.
The observed fluxes of the identified H$_2$ lines are
listed in Table~\ref{h2_table}.   The spectrum is rich also in H~{\sc i}
recombination lines and atomic and ionic 
fine structure lines.  Between 4.5 to 5~$\mu$m, we find
a forest of gaseous 1-0 ro-vibrational CO emission, possibly mixed
with absorption of solid CO (van Dishoeck et al. \cite{dis98}).
Gaseous water is seen in emission through the $\nu_2$ bending mode between
6.3 and 7~$\mu$m and several lines between 30 and 45~$\mu$m, and
gaseous $\rm CO_2$ is detected at 15~$\mu$m.  
PAH features dominate the emission
between 6 and 12~$\rm \mu m$.  Absorption features of water ice 
are seen at 3.1~$\mu$m, of $\rm CO_2$
ice at 4.25~$\rm \mu m$, and of silicate at 9.7~$\rm \mu m$.

The various observed lines and features probe different regions -- both
within the SWS aperture and along the line of sight. The H~{\sc ii} region in
the foreground of OMC-1 contributes to the H recombination and ionic
fine structure emission, whereas the PAH (Verstraete et al. \cite{ver96})
emission and a large fraction of
the [Si~{\sc ii}]34.8$\rm \mu m$ emission (Haas et al.
\cite{haa91}) originate in the PDR between the H~{\sc ii} region
and the molecular cloud which embeds OMC-1. The PDR also contributes to
the $\rm H_2$ emission.  The shocks dominate the emission of H$_2$,
H$_2$O and CO, and may make a minor contribution to the H recombination and most
fine structure emission.


\subsection{Observed H$_2$ level column densities}

All molecular hydrogen lines, due to the small radiative transition
probabilities, remain optically thin.  Therefore the corresponding
``observed'' upper level column density can be computed from the
observed line flux,
\begin{equation} \label{eq:col}
N_{\rm obs}(v,J) = {4\pi \lambda \over h c} 
  {I_{\rm obs}(v,J\rightarrow
  v',J')\over A(v,J\rightarrow v',J')}~,
\end{equation}
where $I_{\rm obs}(v,J\rightarrow v',J')$ and $A(v,J\rightarrow v',J')$
are the observed line flux and the Einstein-$A$ radiative transition
probability of the transition from level $(v,J)$ to $(v',J')$,
respectively. The Einstein coefficients are adopted from Turner et al.
(\cite{tur77}) and Wolniewicz et al. (\cite{wol98}). The transition
energies we computed from level energies kindly provided by E. Roueff
(1992, private communication).

\begin{figure}[htb] 
\begin{center}
\includegraphics[width=1.0\columnwidth]{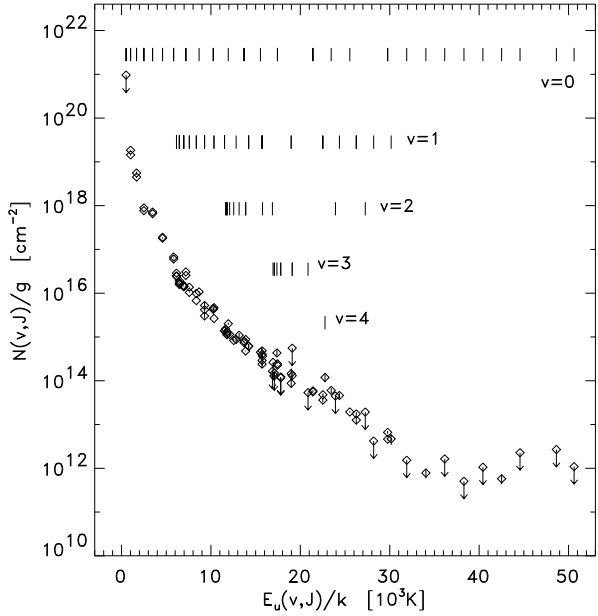}
\caption{Excitation diagram of the H$_2$ level column density
  distribution toward Peak 1, plotting the observed level columns (not
  corrected for extinction, divided by the level degeneracy) against
  level energy.}
\label{fig:boltz1}
\end{center}
\end{figure}

A convenient way to visualize the level column densities is 
to divide them by the level degeneracy $g_J$, and plot this
against the upper level
energy $E_{\rm u}(v,J)/k$; the degeneracy $g_J \equiv g_s (2 J + 1)$,
where $g_s = 3$ for ortho (odd $J$) H$_2$ and $g_s = 1$ for para (even
$J$) H$_2$.  For the lines we observed toward Peak 1 we found that in such
a ``Boltzmann diagram'' Fig.~\ref{fig:boltz1} the level columns 
show a smooth distribution, where the level columns line
up irrespective of their quantum numbers.  There is no sign
of fluorescent excitation or of a deviation from the
ortho-to-para H$_2$ ratio of three.

Fluorescently excited $\rm H_2$, as seen in photodissociation regions
(Timmermann et al. \cite{tim96}) would produce a level distribution in
which the ``rotational temperature'' derived from levels at given $v$ is
lower than the ``vibrational temperature'' derived from levels of
the same $J$ (e.g. Draine \& Bertoldi \cite{dra96}). Fluorescent
excitation therefore shows a characteristic jigsaw distribution of the
$v>1$ levels, unlike the smooth line-up we observed here, where $N/g$
appears not to depend on the state quantum number. Furthermore,
fluorescently excited gas usually shows ortho-to-para ratios in
vibrationally excited levels smaller than the total ortho-to-para ratio
of the gas along the line of sight.  This is due to the enhanced
self-shielding, and therefore reduced excitation rate, of the more
abundant ortho-H$_2$ (Sternberg \& Neufeld \cite{ste99}).

\subsection{Extinction} \label{ex}

The shocks emitting the strong infrared lines in the OMC-1 outflow are
deeply embedded in the molecular cloud, so that the emerging radiation
suffers significant extinction. To correct the column densities derived
from the H$_2$ emission line intensities, $N_{\rm obs}(v,J)$, for this
extinction, we need to know the proper extinction correction as a
function of wavelength.  However, this interstellar infrared extinction
curve is not well determined. Especially the depth and
width of the silicate absorption features, centered at 9.7~$\rm \mu$m
and 18~${\rm \mu m}$ are uncertain, and they could vary from region to
region (Draine \cite{dra89}). A further complication arises from the
mixing of the emitting and absorbing gas, which we might expect in the
outflow regions considering the complex spatial variation of the near-IR
emission mapped in OMC-1 (Fig.~\ref{schultz}), or in similar outflows
such as DR21 and Cep A (Davis \& Smith \cite{dav96}; Goetz et al.
\cite{goe98}).

With enough redundancy in the information provided by the molecular and
atomic lines in Peak~1, we are in principle 
able to estimate the average extinction
along our line of sight as a function of wavelength.  However, the
H~{\sc i} recombination lines and the H$_2$ emission lines may not be
tracing the same regions, and might therefore be subject to differing
extinction. We therefore treat them separately.  To correct the $\rm
H_2$ line fluxes for extinction, we tried to derive the extinction from
the $\rm H_2$ lines directly (Bertoldi et al. \cite{ber99}).

\begin{figure}[htb] 
\begin{center}
\includegraphics[width=1.0\columnwidth]{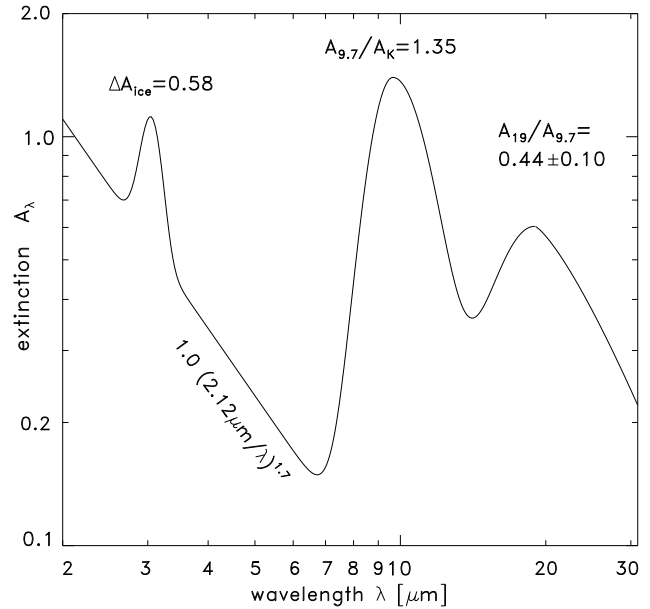}
\caption{Near- and mid-infrared extinction (Eq.~2) found from the relative
  intensities of the H$_2$ lines observed toward Peak~1. The curve was
  constructed using four free parameters for which values 
  were derived that minimize the
  dispersion of the H$_2$ column density distribution (Fig.~5) for
  levels with $E/k < 16\, 000$~K}
\label{extinction}
\end{center}
\end{figure}

An inspection of the excitation diagram Fig.~\ref{fig:boltz1} derived
from the line intensities ({\em uncorrected} for extinction) shows that
the column densities follow a smooth distribution, with no dependence on
vibrational quantum number, and no sign of an ortho-to-para column
density ratio different from three.  Transitions from a given state to
different lower states produce lines with different wavelengths which
suffer extinction.  Deviations from the expected
line ratios of lines from the same state therefore yield the difference
in extinction between the corresponding wavelengths of the lines. More
generally, we can use this to estimate the extinction as a function of
wavelength for a large set of lines, by minimizing the dispersion in the
excitation diagram around a least-squares fit to the level columns.  We
thereby assume that the dispersion in the column densities is
partly due to extinction.  

We constructed an extinction curve (Fig.  \ref{extinction}) with four
free parameters: the absolute normalization for a $A_\lambda\propto\lambda^{\rm
  -1.7}$ power law extinction curve from $\rm 2.4\, \mu m$ to $6\rm \,
\mu m$, the width and depth of the water ice absorption feature at
3.1$\,\mu$m, and the depth of the $9.7\, \mu$m silicate absorption
feature.  We fixed the width of the 9.7 and $\rm 18 \, \mu m$ silicate
features from calculations by Draine \& Lee (\cite{dra84}). The depth of
the 18$\, \mu$m feature was taken to be 0.44 times that of the 9.7$\rm
\,\mu m $ feature, based on an average of previous estimates (Draine \&
Lee \cite{dra84}; Pegourie \& Papoular \cite{peg85}; Volk \& Kwok
\cite{vol88}; Bertoldi et al. in prep.). 
Calibration uncertainties and a small contribution of foreground
fluorescently excited H$_2$ may give rise to a dispersion in the
derived extinction corrections that is not due to extinction.
The derived  curve should therefore be considered as very approximate. 
The extinction curve minimizing the dispersion in the excitation curve
is shown in Fig.~\ref{extinction}.  Explicitly it can be written
\begin{eqnarray}
  A(\lambda) & = &  A_{\rm K} (\lambda/2.12)^{-1.7} + 0.58~e^{-22(\lambda-3.05)^2} \nonumber \\
             &   &  {} + (1.35-0.08 A_{\rm K}) \left\{e^{-[c_1 \log(\lambda/9.66)]^2} \right. \\
             &   &  \left. {} + 0.44 ~e^{-[c_2 \log(\lambda/19)]^2} \right\}~, \nonumber             
\end{eqnarray}
where $A_{\rm K}=(1.0\pm 0.1)$~mag is the implied extinction at 2.12~\mum, and
$c_1=14.3$ for $\lambda<9.7$,
$c_1=9.8$ for  $\lambda>9.7$,
$c_2=7.5$ for  $\lambda<19$,
$c_2=4.8$ for  $\lambda>19$, and $\lambda$ is given here in \mum.
The depth of the extinction minimum at 6.5~$\mu$m is very uncertain,
since it is not constrained by the inconsistent corrections derived from
the four lines between 5.8 and 7.3~$\mu$m. There is an indication
that the minimum is at least as deep as our simple curve 
shows, but a much more
careful analysis of the line fluxes would be necessary to reach a firm
conclusion.
  
Atomic hydrogen recombination lines could offer another means to trace the
extinction as a function of wavelength. We detected seventeen H~{\sc i}
recombination lines ranging in wavelength from 2.6 to 19~$\rm \mu m$.
Since the relative emissivities are known from theory (Storey \& Hummer
\cite{sto95}) and depend only mildly on the gas temperature and density,
a comparison of the observed line intensities divided by their
respective case B emissivities yields a measure for the differential
extinction between the respective lines' wavelengths.

\begin{figure*}[htb] 
\begin{center}
\includegraphics[width=1.5\columnwidth]{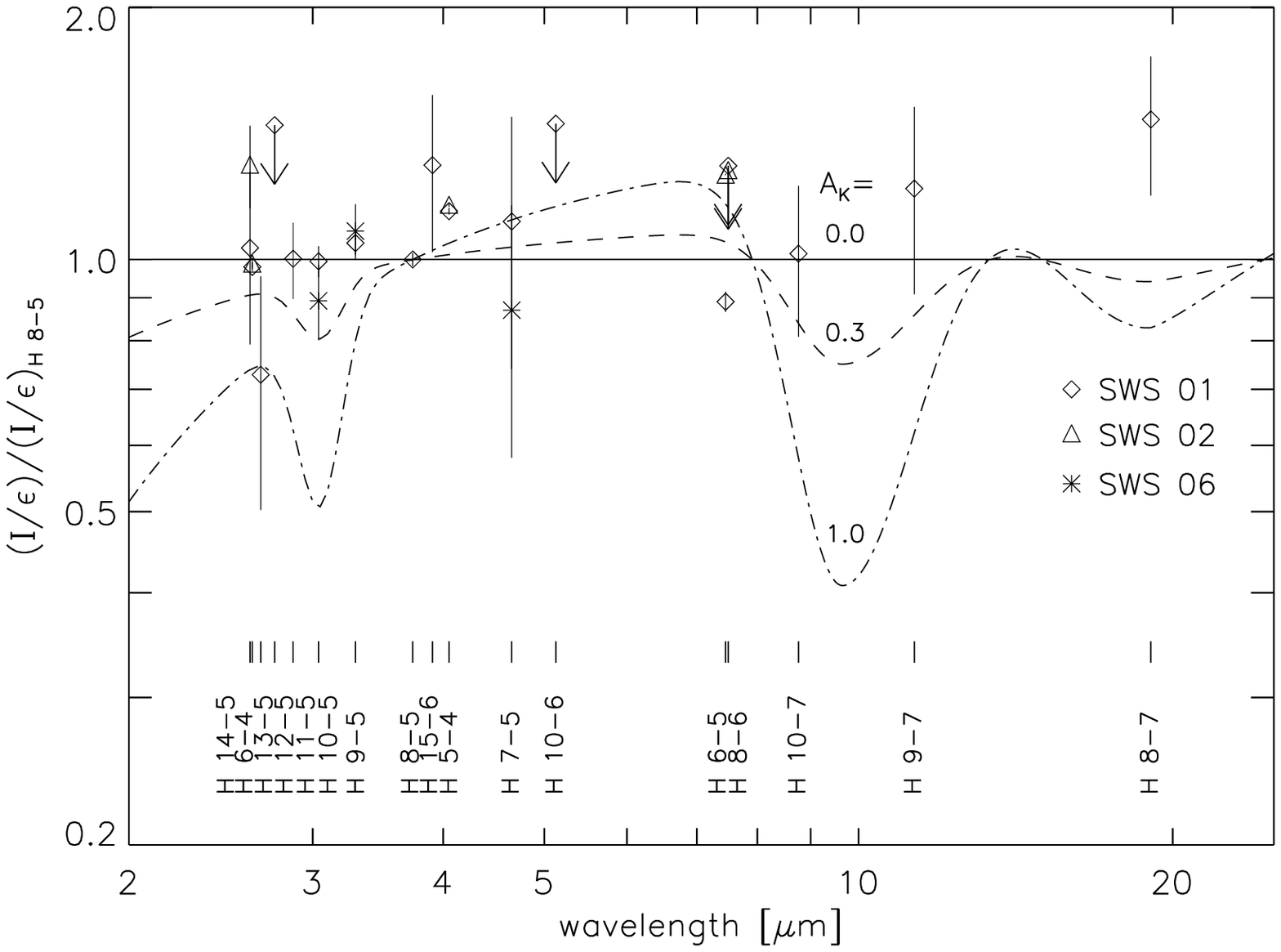}
\caption{An attempt to derive the differential extinction 
  in the H recombination line emission:  The line intensities divided by
  their respective emissivities are normalized by this value for the H
  8--5 line, and then plotted against line wavelength.  With no
  differential extinction to the H 8--5 line, all data points should
  line up at unity value.  The broken lines show the expected
  distribution of values adopting the shape of the extinction curve
  derived from H$_2$ lines (Fig.\ref{extinction}), with two different
  values of the absolute extinction at K.  The apparent distribution is
  consistent with an extinction of zero.}
\label{recom}
\end{center}
\end{figure*}

In Fig.~\ref{recom} we plot against wavelength the observed line
intensities, divided by their emissivities and normalized to this ratio
for the H~{\sc i} 8--5 transition line. The data points scatter around
unity, which means that there is little if any differential extinction
over this wavelength range.  A comparison with the distribution of
intensities expected for an extinction curve with the shape we found
from the H$_2$ lines reveals none of the prominent extinction features.
A reasonable explanation is that the total extinction to the H~{\sc i}
emission region is very low, $A_{\rm K}<0.3$ mag.

Although the errors are too large to constrain the exact value of the
extinction, it is obvious that the H recombination lines are much
less attenuated than the H$_2$ lines.  This suggests that the bulk of
the atomic hydrogen emission arises in the foreground H~{\sc ii} region,
whereas the $\rm H_2$ emitting region is more deeply embedded
in the molecular cloud.

This conclusion agrees with the assessment of Everett et al.
(\cite{eve95}), who obtained $A_{\rm J}=(0.38 \pm 0.09)$~mag for the
extinction shown by H recombination lines, but found $A_{\rm J}=(2.15
\pm 0.26)$~mag from the H$_2$ lines. With a $\lambda^{-1.7}$ extinction
law this corresponds to $A_{\rm K}=(0.15 \pm 0.04)$~mag and $A_{\rm
  K}=(0.9 \pm 0.1)$~mag, in good agreement with our results.

\subsection{Fine structure lines} \label{finestructure}

The observed atomic and ionic fine structure lines are valuable
diagnostics. If highly-ionized species are found toward the shocked
region -- which is well shielded from the ionizing radiation of the
Trapezium stars -- they would indicate the presence of fast, ionizing
J-shocks.  It is therefore interesting to disentangle the respective
contributions to the fine structure line emission of the foreground
H~{\sc ii} region/PDR and the shocked gas of the OMC-1 outflow.

Table~\ref{fine} lists the observed intensities of a number of
fine structure lines we searched for toward Peak 1.

\begin{table*}
\caption[]{
ISO-SWS Observations of Fine Structure Lines toward Orion Peak~1
}
\begin{tabular}{lrrrrccrlc}
\hline\noalign{\smallskip}
\hline\noalign{\smallskip}
\multicolumn{1}{c}{ } & 
\multicolumn{1}{c}{$\lambda$} & 
\multicolumn{1}{c}{$\rm IP_l~^a$} & 
\multicolumn{1}{c}{$\rm IP_u~^b$} &
\multicolumn{1}{c}{$n_{\rm crit}~^{\rm c}$} &
\multicolumn{1}{c}{SWS} &
\multicolumn{1}{c}{$I_{\rm obs}$} &
\multicolumn{1}{c}{S/N~$^{\rm d}$ } &
\multicolumn{1}{c}{$A_\lambda~^{\rm e}$} &
\multicolumn{1}{c}{$I_{\rm model}~^{\rm f}$} \\
\multicolumn{1}{c}{} & 
\multicolumn{1}{c}{[$\rm \mu$m]}    & 
\multicolumn{1}{c}{[eV]}        &      
\multicolumn{1}{c}{[eV]} & 
\multicolumn{1}{c}{[cm$^{\rm -3}$]} &
\multicolumn{1}{c}{AOT} &
\multicolumn{1}{c}{[erg~s$^{\rm -1}$cm$^{\rm -2}$sr$^{\rm -1}$]} &
\multicolumn{1}{c}{} &
\multicolumn{1}{c}{[mag]} &
\multicolumn{1}{c}{[erg~s$^{\rm -1}$cm$^{\rm -2}$sr$^{\rm -1}$]} \\
\noalign{\smallskip}
\hline\noalign{\smallskip}
[Ne~{\sc iii}] & 36.009 & 40.96 & 63.45 & $5.52 \times 10^4$ & 01 & $2.26 \times 10^{-3}$ & 4.3  & 0.02 & $1.78 \times 10^{-3}$ \\
{[Fe~{\sc ii}]}  & 35.777 & 7.90  & 16.19 &                    & 01 & $1.22 \times 10^{-3}$ & 2.0  & 0.02 & \\
{[Si~{\sc ii}]}  & 34.814 & 8.15  & 16.35 & $3.67 \times 10^{\rm 5~g}$ & 01 & $1.43 \times 10^{-2}$ & 19.1 & 0.03 & $3.16 \times 10^{-4}$ \\
               &        &       &       &                         & 02 & $1.59 \times 10^{-2}$ & 58.8 &        & \\
{[S~{\sc iii}]}  & 33.480 & 22.34 & 34.79 & $6.34 \times 10^3$ & 01 & $2.16 \times 10^{-2}$ & 46.0 & 0.03 & $1.20 \times 10^{\rm -2~h}$\\
{[Fe~{\sc ii}]}  & 25.988 & 7.90  & 16.19 & $3.59 \times 10^{\rm 4~i}$ & 01 & $3.70 \times 10^{-3}$ & 6.8  & 0.05 & \\
               &        &       &       &                           & 02 & $1.76 \times 10^{-3}$ &10.6  &      &  \\
{[S~{\sc i}]}    & 25.249 & 0.0   & 10.36 & $4.2 \times 10^{\rm 4~j}$  & 01 &$1.19 \times 10^{-2}$ & 19.5 & 0.35$~^{\rm k}$ & \\
               &        &       &       &                           & 02 & $1.17 \times 10^{-2}$ & 20.5 &   &  \\
{[Fe~{\sc iii}]} & 22.925 & 16.19 & 30.65 & $1.12 \times 10^5$ & 01 & $4.15 \times 10^{-4}$ & 2.1  & 0.07 & \\
{[S~{\sc iii}]}  & 18.713 & 23.34 & 34.79 & $2.06 \times 10^4$ & 01 & $3.61 \times 10^{-2}$ & 307.0 & 0.09 & $3.63 \times 10^{-2}$\\
{[Fe~{\sc ii}]}  & 17.936 & 7.90  & 16.19 &                    & 01 & $2.56 \times 10^{-4}$ & 2.2  & 0.09 & \\
{[P~{\sc iii}]}  & 17.885 & 19.77 & 30.20 & $3.91 \times 10^4$ & 01 & $3.09 \times 10^{-4}$ & 2.9  & 0.09 & \\
{[Ne~{\sc iii}]} & 15.555 & 40.96 & 63.45 & $2.70 \times 10^5$ & 01 & $3.20 \times 10^{-2}$ & 255.0 & 0.07 & $2.80 \times 10^{-2}$\\
{[Ne~{\sc ii}]}  & 12.814 & 21.56 & 40.96 & $6.54 \times 10^5$ & 01 & $3.03 \times 10^{-2}$ & 109.0 & 0.07 & $3.16 \times 10^{-2}$ \\
               &        &       &       &                    & 02 & $3.11 \times 10^{-2}$ & 63.3 &       &  \\
{[S~{\sc iv}]}$^{\rm l}$ & 10.511 & 34.79 & 47.22 & $5.39 \times 10^4$ & 01 & $9.69 \times 10^{-3}$ & 550.0 & 0.19 & $2.40 \times 10^{-2}$ \\
{[Ar~{\sc iii}]} & 8.991  & 27.63 & 40.74 & $3.18 \times 10^5$ & 01 & $9.29 \times 10^{-3}$ & 323.0 & 0.18 & $1.38 \times 10^{-2}$\\
{[Ar~{\sc ii}]}  & 6.985  & 15.76 & 27.63 & $4.17 \times 10^5$ & 01 & $2.92 \times 10^{-3}$ & 148.0 & 0.02 & $2.82 \times 10^{-3}$\\
{[Ni~{\sc ii}]}$^{\rm m}$ & 6.636  & 7.64  & 18.17 & $1.73 \times 10^7$ & 01 & $1.26 \times 10^{-3}$ &  47.8 & 0.02 & \\
\noalign{\smallskip}        \hline
\noalign{\smallskip}
\end{tabular}

$^{\rm a}$ Lower ionization potential, IP$_{\rm l}$, to produce the ion.\\
$^{\rm b}$ Upper ionization potential, IP$_{\rm u}$, of the next higher ionization stage.\\
$^{\rm c}$ Collisions with electrons at $T = 10\,000$~K, unless indicated.\\
$^{\rm d}$ Calculated from the RMS noise within $\rm \sim 500 \: km\,s^{-1}$. \\
$^{\rm e}$ From the extinction law of
Bertoldi et al. (\cite{ber99}) with $A_{\rm K}=$0.15 mag.\\
$^{\rm f}$ Best model by Rubin et al. (\cite{rub91}) for a projected
distance of $\sim 86\arcsec$ from $\theta^1$~Ori~C. \\ 
$^{\rm g}$ For collisions with electrons at $20\,000$~K. Collisions with
H atoms at 300 K yield $n_{\rm crit} = 3.67 \times 10^5 \rm\: cm^{-3}$.\\
$^{\rm h}$ From the predicted [S~{\sc iii}]18.7$\mu$m line intensity and the
[S~{\sc iii}]18.7$\mu$m/[S~{\sc iii}]33.5$\mu$m line ratio. \\
$^{\rm i}$ Collisions with H atoms yield $n_{\rm crit} = 2.24 \times 10^6$ cm$^{-3}$.\\
$^{\rm j}$ For collisions with $\rm H^+$ ions. Collisions with
H atoms at 300 K yield $n_{\rm crit} = 1.5 \times 10^6 \rm\: cm^{-3}$.\\
$^{\rm k}$ Because [S~{\sc i}]25.249$\rm \mu m$ probably arises from
the more deeply embedded, shocked region, we adopt  $A_{\rm K} = 1.0$ as
for the H$_2$ emission, with the
extinction curve of Fig.~\ref{extinction}.\\
$^{\rm l}$ Merged with H~12-8, but we estimate latter to
contribute only $4\times 10^{-5} \: {\rm erg~s^{-1}cm^{-2}sr^{-1}}$, adopting case-B
emissivities. \\
$^{\rm m}$ Merged with an $\rm H_2O$ line at 6.6354 $\rm \mu m$.\\

\label{fine}
\end{table*} 


A predominantly ionized medium is traced by species with ionization
potentials larger than 13.6 eV.  From such ions a number of lines,
[Ar~{\sc ii}]6.9$\mu$m, [Ar~{\sc ii}]8.99$\mu$m, [Ne~{\sc
  ii}]12.8$\mu$m, [Ne~{\sc iii}]15.5$\mu$m, [Ne~{\sc iii}]36$\mu$m,
[S~{\sc iii}]18.7$\mu$m, and [S~{\sc iv}]10.5$\mu$m, are found in our
ISO-SWS spectra, and their intensities can be compared with H~{\sc ii}
region models such as those computed by Rubin et al. (\cite{rub91}).
A comparison of the line intensities and their ratios to a blister H~{\sc ii}
region model with a star of $T_{\rm eff}=37\,000$~K and $\log g = 4.0$, shows
good agreement of all line intensities, except for that the models
overestimate the [S~{\sc iii}]18.7$\mu$m/[S~{\sc iii}]33.5$\mu$m ratio
by a factor 1.8.  The good agreement indicates that these lines may be
predominantly produced in the foreground H~{\sc ii} region, although a
shock contribution of up to 30\% cannot be excluded.

We can compare the Peak 1 fine structure line emission also with that
seen toward the Orion Bar photodissociation region and ionization front,
which is also irradiated by the Trapezium stars.  We know that here no
fast shocks should contribute to the emission, and that the emission
should be similar to that coming from the PDR in front of the OMC-1
outflow (Herrmann et al. \cite{her97}).  In Table~\ref{bar} we list line
intensities we observed toward two positions on the Orion Bar: toward
the ionization front at $5^{\rm h} 35^{\rm m} 19.31^{\rm s}$, $-5 \degr
24 \arcmin 59.9\arcsec$ (J2000), and toward the peak of the H$_2$
1-0 S(1) emission at $5^{\rm h} 35^{\rm m} 20.31^{\rm s}$, $-5 \degr 25
\arcmin 19.9\arcsec$.  A comparison with the Peak 1 intensities shows
that the intensities of most lines agree within a factor of a few,
suggesting that ionic emission indeed arises in the foreground H~{\sc ii} 
region.

The [P~{\sc iii}]17.9$\mu$m, [Fe~{\sc iii}]22.9$\mu$m, [Fe~{\sc
  ii}]26$\mu$m, and the [S~{\sc iii}]33.5$\mu$m lines were not included
in the Rubin models, but their intensities are very similar toward the
outflow and the Bar.  It is unclear where the [Fe~{\sc ii}]26$\mu$m
emission comes from, though.  It could be produced either in the PDR or
in the ionization front.  A detailed analysis of the [Fe~{\sc ii}]
emission will be subject of a subsequent publication (Bertoldi et al.,
in prep.).

\begin{table*}
\caption[]{
Fine Structure Lines toward two Positions on the Orion Bar PDR
}
\begin{tabular}{lrccrcrccrc}
\hline\noalign{\smallskip}
\hline\noalign{\smallskip}
\multicolumn{2}{c}{ } & 
\multicolumn{4}{c}{\rule[-2mm]{0mm}{6mm}Bar H$_2$ S(1)} & 
\multicolumn{1}{c}{ } &
\multicolumn{4}{c}{Bar Br$\gamma$} \\ 
\cline{3-6}
\cline{8-11}
\multicolumn{1}{c}{ } & 
\multicolumn{1}{c}{$\lambda$} & 
\multicolumn{1}{c}{SWS} &
\multicolumn{1}{c}{$I_{\rm obs}$} &
\multicolumn{1}{c}{ } &
\multicolumn{1}{c}{ } &
\multicolumn{1}{c}{ } &
\multicolumn{1}{c}{SWS} &
\multicolumn{1}{c}{$I_{\rm obs}$} &
\multicolumn{1}{c}{ } &
\multicolumn{1}{c}{ } \\
\multicolumn{1}{c}{} &
\multicolumn{1}{c}{[$\rm \mu$m]}    & 
\multicolumn{1}{c}{AOT} &
\multicolumn{1}{c}{[erg~s$^{\rm -1}$cm$^{\rm -2}$sr$^{\rm -1}$]} &
\multicolumn{1}{c}{\raisebox{1.5ex}[-1.5ex]{S/N$~^{\rm a}$}} &
\multicolumn{1}{c}{\raisebox{1.5ex}[-1.5ex]{Bar/Pk1 $~^{\rm b}$}} &
\multicolumn{1}{c}{ } &
\multicolumn{1}{c}{AOT} &
\multicolumn{1}{c}{[erg~s$^{\rm -1}$cm$^{\rm -2}$sr$^{\rm -1}$]} &
\multicolumn{1}{c}{\raisebox{1.5ex}[-1.5ex]{S/N}} & 
\multicolumn{1}{c}{\raisebox{1.5ex}[-1.5ex]{Bar/Pk1 $~^{\rm b}$}} \\
\noalign{\smallskip}
\hline\noalign{\smallskip}
{[Ne~{\sc iii}]} & 36.009 & 01 & $1.75 \times 10^{-3}$ &  12.9 &  0.8  & & 01 & $1.31 \times 10^{-3}$ & 1.8  & 0.6 \\
{[Fe~{\sc ii}]}  & 35.777 & 01 & $2.05 \times 10^{-4}$ &   2.8 &  0.2  & &        &                       &      &     \\
{[Si~{\sc ii}]}  & 34.814 & 01 & $1.38 \times 10^{-2}$ & 135.0 &  1.1  & & 01 & $1.16 \times 10^{-2}$ & 7.1  & 1.0 \\
               &        &        &                       &       &       & & 02 & $1.13 \times 10^{-2}$ & 63.7 & 0.7 \\
{[S~{\sc iii}]}  & 33.480 & 01 & $2.70 \times 10^{-2}$ &  88.4 &  1.3  & & 01 & $4.78 \times 10^{-2}$ & 36.4 & 2.2 \\
{[Fe~{\sc ii}]}  & 25.988 & 01 & $1.38 \times 10^{-3}$ &  19.4 &  0.4  & & 01 & $9.58 \times 10^{-4}$ &  4.2 & 0.3 \\
  &      &        &                       &       &       & & 02 & $1.37 \times 10^{-3}$ & 17.8 & 0.8 \\
{[S~{\sc i}]}$^{\rm d}$  & 25.249 & 01 & $ < 1.9 \times 10^{-4}$ &   &$ < 0.02$ & & 01 & $< 2.1 \times 10^{-4}$ &  &$< 0.02$  \\                                               
{[Fe~{\sc iii}]} & 22.925 & 01 & $4.82 \times 10^{-4}$ &  10.6 &  1.2  & & 01 & $1.64 \times 10^{-3}$ &  4.4 & 4.0 \\
               &        &        &                       &       &       & & 02 & $9.60 \times 10^{-4}$ & 15.6 & 2.3 \\
{[Ar~{\sc iii}]} & 21.842 & 01 & $3.59 \times 10^{-4}$ &   6.0 &       & &        &                       &      &     \\
{[S~{\sc iii}]}  & 18.713 & 01 & $2.08 \times 10^{-2}$ & 391.0 &  0.6  & & 01 & $5.19 \times 10^{-2}$  & 249.0 & 1.4 \\  
{[Fe~{\sc ii}]}  & 17.936 & 01 & $4.22 \times 10^{-5}$ &   2.1 &  0.1  & &        &                       &      &     \\
{[P~{\sc iii}]}  & 17.885 & 01 & $1.56 \times 10^{-4}$ &   6.6 &  0.5  & &        &                       &      &     \\
{[Ne~{\sc iii}]} & 15.555 & 01 & $7.54 \times 10^{-3}$ & 186.0 &  0.2  & & 01 & $2.62 \times 10^{-2}$ & 575.0 & 0.8 \\
{[Ne~{\sc ii}]}  & 12.814 & 01 & $1.49 \times 10^{-2}$ &  76.5 &  0.5  & & 01 & $3.44 \times 10^{-2}$ & 70.6 & 1.1 \\
{[S~{\sc iv}]}$^{\rm c}$ & 10.511 & 01 & $2.29 \times 10^{-3}$ & 100.0 & 0.2   & & 01 & $8.38 \times 10^{-3}$ & 55.0 & 0.9 \\
{[Ar~{\sc iii}]} &  8.991 & 01 & $3.69 \times 10^{-3}$ & 116.0 &  0.4  & & 01 & $1.13 \times 10^{-2}$ & 194.0 & 1.2 \\
{[Ar~{\sc ii}]}  & 6.985  & 01 & $1.90 \times 10^{-3}$ &  18.7 &  0.7  & & 01 & $1.15 \times 10^{-2}$ &  40.2 & 3.9 \\
{[Ni~{\sc ii}]}$^{\rm d}$  & 6.636  & 01 & $< 7.7 \times 10^{-5}$ & & $< 0.06$ & & 01 & $< 1.4 \times 10^{-4}$ &   & $< 0.1$ \\          
\noalign{\smallskip}        \hline
\noalign{\smallskip}
\end{tabular}

$^{\rm a}$ From the RMS noise within $\rm \sim 500 \: km\,s^{-1}$. \\
$^{\rm b}$ Ratio of the respective intensities toward Bar H$_2$ S(1) and
Peak 1, and Bar Br$\gamma$ and Peak 1. If lines in different observing
modes are available the ratio was calculated from  lines of the same mode.\\
$^{\rm c}$ The [S~{\sc iv}] line is merged with the H~12-8 line.\\
$^{\rm d}$ The upper limit for the intensities is calculated from the $3 \sigma$ flux density noise level at
the respective wavelength times the width of one resolution element.

\label{bar}
\end{table*} 


\paragraph{Silicon:}
Haas et al. (\cite{haa91}) observed [Si~{\sc ii}]34.8$\mu$m strip maps
across the OMC-1 outflow.  From the apparent peak of emission near IRc2
they concluded that about half of this emission must be due to the
production and excitation of gas phase silicon in shocks.  In their
preliminary reduction of a $\sim 6 \arcmin$ square map of [Si~{\sc ii}],
Stacey et al. (\cite{sta95}) also find that the emission peaks toward
the OMC-1 outflow, and this excess is consistent with that observed by
Haas et al. (\cite{haa91}).  Haas et al. find a surface flux density of
$6\times 10^{-3}\rm ~erg~cm^{-2}s^{-1}sr^{-1}$ toward Peak 1, of which
they attribute about half to an extended component, which most likely
arises from the PDR lining the foreground H~{\sc ii} region.  Haas et
al. find that the flux is similar toward Peak~1 and the Orion Bar, which
may well be due to limb brightening by a factor two of the PDR component
at the Bar.  Our observations of [Si~{\sc ii}]34.8$\mu$m toward the Bar
and Peak 1 yield fluxes twice as high as those of Haas.  This could be
due either to a calibration error, or to beam dilution in the Haas et
al. measurements. Either way, it seems that both the PDR and the shocks
give rise to strong silicon emission, which for the PDR at least
requires Si gas phase abundances of order 10\% solar: the PDR models by
Tielens \& Hollenbach (\cite{tie85b}) adopt a 2.2\% solar gas phase Si
abundance and predict about a quarter of the flux we could attribute to
the PDR.  The silicon abundance can be enhanced in shocks by sputtering
(Martin-Pintado et al. \cite{mar92}; Caselli et al. \cite{cas97};
Bachiller \& Perez-Gutierrez \cite{bac97}). Large gas phase silicon
abundances are also found in other PDRs such as NGC 7023 (Fuente et al.
\cite{fue99}). The mechanism by which the abundance is enhanced in
PDRs is still unclear, although photodesorption has been suggested
(Walmsley et al. \cite{wal99}). Strong silicate emission is however not
a universal feature of PDRs: based on ISO observations of 
[Si~{\sc ii}]34.8$\mu$m toward NGC 2023 and a comparison with model
calculations, Draine \& Bertoldi (2000) report Si to be quite highly
depleted in the NGC 2023 PDR.

\paragraph{Other lines:}
Of the other fine structure lines seen toward Peak 1, [Ni~{\sc
  ii}]6.6$\mu$m and [S~{\sc i}]25$\mu$m are not detected toward the
Orion Bar. The [Ni~{\sc ii}]6.6$\mu$m line is confused with a water
line, making it difficult to detect.  Although the [Fe~{\sc ii}]18$\mu$m
and [Fe~{\sc ii}]36$\mu$m lines are marginally detected in both objects,
they both appear to be an order of magnitude fainter toward the Bar than
toward Peak~1. This suggests that shocks are more efficient in producing
and exciting gas phase iron.

\paragraph{Sulfur:}
The strong [S~{\sc i}]25$\mu$m line emission is probably shock-excited.
Burton et al. (\cite{bur90a}) computed the [S~{\sc i}] intensity in
their PDR model for densities of $10^3$ to $\rm 10^5 \: cm^{-3}$ and
radiation fields of $10^3$ to $10^5$ times the ambient interstellar
field as $\rm \leq 10^{-5} \: erg \, s^{-1} \, cm^{-2} \, sr^{-1}$,
which is three orders of magnitude below our observed [S~{\sc i}]
intensity. Both a J- or C-type shock could account for the [S~{\sc i}]
emission. But 
only a J-shock is able to produce both the [S~{\sc i}] and the 
[Si~{\sc ii}] line emission.

We compared the estimated shock contribution to the observed [Si~{\sc
  ii}]34.8$\mu$m flux of $\rm \sim 7 \times 10^{-3} $ $\rm erg \, s^{-1}
\, cm^{-2} \, sr^{-1}$ (Haas et al. \cite{haa91}) and the [S~{\sc
  i}]25$\mu$m line flux to the J-shock model of Hollenbach \& McKee
(\cite{hol89}).
Both the relative and absolute [S~{\sc i}] and [Si~{\sc ii}] fluxes could
be explained by shocks of high velocities, $v_{\rm s} = {\rm (85 \pm 10)
  \: km\,s^{-1}}$, a pre-shock hydrogen nuclei density $ n_{\rm H} = {\rm
  (10^5 - 10^6) \: cm^{-3}}$, and a beam filling factor $\phi \sim 3-4$.
A beam-filling planar shock results in $\phi = 1$, and a 
beam-filling spherical shock in $\phi = 4$.  A shock contribution of 10 to
30\% to the [Ne~{\sc ii}]12.8$\mu$m flux would also explain the observed
[Ne~{\sc ii}]/[Si~{\sc ii}] and [Ne~{\sc ii}]/[S~{\sc i}] flux ratios.


\subsection{Molecular hydrogen} \label{molh2}

In the spectra shown in Figs.~\ref{spectrum}, \ref{spec}, and
\ref{multi}, we detected 56 different $\rm H_2$ lines of pure rotational
and rotation-vibrational transitions (Table~\ref{h2_table}).  Pure
rotational lines were detected ranging from the 0-0~S(1) to 0-0~S(25)
transitions, which correspond to upper level energies $E(v,J)/k$ ranging
from 1015~K to $42\,500$~K. Adding a large number of
vibration-rotational transition lines, we are able to study the
excitation of the gas within the ISO aperture over an unprecedented
range.

The H$_2$ 0-0 S(0) transition line was not detected from our
observations with the medium resolution grating modes (SWS~01 and
SWS~02, $R \sim 1000-2000$).  Unfortunately, our observation with the
Fabry-Perot did not cover a spectral range wide enough to detect a line
with the expected width of $\sim 60\rm ~km ~s^{-1}$ (Nadeau \& Geballe
\cite{nad79}; Brand et al. \cite{bra89b}, Moorhouse et al. \cite{moo90};
Chrysostomou et al. \cite{chr97}).  However, the FP spectrum, shown in
Fig.~\ref{multi}, shows a line-like feature with a a narrow width of
12~km~s$^{-1}$, comparable to the spectral resolution in this observing
mode.  This feature could be emission arising in the
foreground photodissociation region bounding the Orion Nebula and the
dense molecular cloud embedding the outflow.

\begin{table*}[h]
\caption[]{
Summary of the ISO-SWS $\rm H_2$ line observations at Orion Peak~1
}
\begin{tabular}{lrrlrcrrrr}
\hline\noalign{\smallskip}
\hline\noalign{\smallskip}
\multicolumn{1}{c}{ } &
\multicolumn{1}{c}{$\lambda$} &
\multicolumn{1}{c}{$E_{\rm u}/k~^{\rm a}$} &
\multicolumn{1}{c}{$A~^{\rm b}$} &
\multicolumn{1}{c}{SWS} &
\multicolumn{1}{c}{$I_{\rm obs}~^{\rm c}$} &
\multicolumn{1}{c}{ } &
\multicolumn{1}{c}{$A_\lambda~^{\rm e}$} &
\multicolumn{1}{c}{$N_{\rm u}~^{\rm f}$} \\
\multicolumn{1}{c}{\raisebox{1.ex}[-1.ex]{line}} &
\multicolumn{1}{c}{[$\rm \mu m$]} &
\multicolumn{1}{c}{[K]} &
\multicolumn{1}{c}{[s$^{-1}$]} &
\multicolumn{1}{c}{AOT} &
\multicolumn{1}{c}{[erg~s$^{\rm -1}$cm$^{\rm -2}$sr$^{\rm -1}$]} &
\multicolumn{1}{c}{\raisebox{1.ex}[-1.ex]{S/N$^{\rm d}$}} &
\multicolumn{1}{c}{[mag]} &
\multicolumn{1}{c}{[cm$^{-2}$]} \\
\noalign{\smallskip}
\hline\noalign{\smallskip}
0-0 S(0)  & 28.2188 &   509.9   & $2.94 \times 10^{-11}$ & 01 & $< 7.90 \times 10^{-4}$ & $< 3$.  & 0.27 & $< 6.17 \times 10^{21}$ \\
0-0 S(1)  & 17.0346 &  1015.0   & $4.76 \times 10^{-10}$ & 01 & $ 1.34 \times 10^{-3}$ &  27.8     & 0.53 & $4.92 \times 10^{20}$  \\
          &         &           &                        & 02 & $ 1.71 \times 10^{-3}$ &  51.4     & 0.53 & $6.28 \times 10^{20}$  \\
0-0 S(2)  & 12.2785 &  1682.0   & $2.76 \times 10^{-9}$  & 01 & $ 1.44 \times 10^{-3}$ &  32.8     & 0.57 & $6.84 \times 10^{19}$  \\
          &         &           &                        & 02 & $ 1.78 \times 10^{-3}$ &  12.6     & 0.57 & $8.46 \times 10^{19}$  \\
0-0 S(3)  &  9.6649 &  2503.4   & $9.84 \times 10^{-9}$  & 01 & $ 4.08 \times 10^{-3}$ & 200.0     & 1.35 & $8.82 \times 10^{19}$  \\
          &         &           &                        & 02 & $ 4.72 \times 10^{-3}$ & 251.0     & 1.35 & $1.02 \times 10^{20}$  \\
0-0 S(4)  &  8.0258 &  3474.6   & $2.64 \times 10^{-8}$  & 01 & $ 4.43 \times 10^{-3}$ & 133.0     & 0.45 & $1.28 \times 10^{19}$  \\
          &         &           &                        & 02 & $ 4.84 \times 10^{-3}$ & 231.0     & 0.45 & $1.40 \times 10^{19}$  \\
1-1 S(5)  &  7.2807 & $10\,340.3$   & $5.44 \times 10^{-8}$  & 01 & $ 2.48 \times 10^{-4}$ &   6.7     & 0.18 & $2.48 \times 10^{17}$  \\
0-0 S(5)  &  6.9091 &  4586.7   & $5.88 \times 10^{-8}$  & 01 & $ 1.09 \times 10^{-2}$ &  63.7     & 0.15 & $9.31 \times 10^{18}$  \\
          &         &           &                        & 02 & $ 1.15 \times 10^{-2}$ &  64.2     & 0.15 & $9.83 \times 10^{18}$  \\
0-0 S(6)  &  6.1089 &  5829.8   & $1.14 \times 10^{-7}$  & 01 & $ 3.37 \times 10^{-3}$ &  44.2     & 0.17 & $1.33 \times 10^{18}$  \\
          &         &           &                        & 02 & $ 3.03 \times 10^{-3}$ &  58.3     & 0.17 & $1.20 \times 10^{18}$  \\
1-1 S(7)  &  5.8111 & $12\,816.4$   & $1.82 \times 10^{-7}$  & 01 & $ 2.48 \times 10^{-4}$ &   8.1     & 0.18 & $5.91 \times 10^{16}$  \\
0-0 S(7)  &  5.5115 &  7196.6   & $2.00 \times 10^{-7}$  & 01 & $ 9.99 \times 10^{-3}$ & 113.0     & 0.20 & $2.09 \times 10^{18}$  \\
          &         &           &                        & 02 & $ 8.33 \times 10^{-3}$ & 141.0     & 0.20 & $1.74 \times 10^{18}$  \\
0-0 S(8)  &  5.0528 &  8677.1   & $3.24 \times 10^{-7}$  & 01 & $ 2.28 \times 10^{-3}$ &  24.8     & 0.23 & $2.78 \times 10^{17}$  \\
1-1 S(9)$^{\rm g}$  &  4.9533 & $15\,725.5$   & $4.38 \times 10^{-7}$  & 01 & $ 4.65 \times 10^{-4}$ &   4.5     & 0.24 & $4.14 \times 10^{16}$  \\
          &         &           &                        & 06 & $ 2.81 \times 10^{-4}$ &   4.1     & 0.24 & $2.50 \times 10^{16}$  \\
0-0 S(9)$^{\rm g}$  &  4.6947 & $10\,261.2$   & $4.90 \times 10^{-7}$  & 01 & $ 5.09 \times 10^{-3}$ &  15.7     & 0.26 & $3.92 \times 10^{17}$  \\ 
          &         &           &                        & 02 & $ 4.85 \times 10^{-3}$ &  26.9     & 0.26 & $3.73 \times 10^{17}$  \\   
1-1 S(11) &  4.4171 & $18\,977.1$   & $8.42 \times 10^{-7}$  & 01 & $ 2.13 \times 10^{-4}$ &   4.6     & 0.29 & $9.21 \times 10^{15}$  \\   
0-0 S(10) &  4.4096 & $11\,940.2$   & $7.03 \times 10^{-7}$  & 01 & $ 1.26 \times 10^{-3}$ &  18.2     & 0.29 & $6.52 \times 10^{16}$  \\   
3-2 O(7)  &  4.3298 & $19\,092.2$   & $9.77 \times 10^{-8}$  & 01 & $< 6.54 \times 10^{-5}$ &  $< 3$     & 0.30 & $< 2.41 \times 10^{16}$  \\   
2-2 S(13) &  4.3137 & $27\,264.3$   & $1.14 \times 10^{-6}$  & 01 & $< 7.47 \times 10^{-5}$ & $< 3$    & 0.30 & $< 2.35 \times 10^{15}$  \\   
0-0 S(11) &  4.1810 & $13\,702.7$   & $9.64 \times 10^{-7}$  & 01 & $ 2.39 \times 10^{-3}$ &  31.7     & 0.32 & $8.77 \times 10^{16}$  \\       
          &         &           &                        & 02 & $ 2.19 \times 10^{-3}$ & 113.0     & 0.32 & $8.03 \times 10^{16}$  \\       
1-0 O(8)  &  4.1622 &  9285.7   & $7.38 \times 10^{-8}$  & 01 & $ 1.91 \times 10^{-4}$ &   3.7     & 0.32 & $9.13 \times 10^{16}$  \\       
          &         &           &                        & 02 & $ 1.10 \times 10^{-4}$ &   7.9     & 0.32 & $5.26 \times 10^{16}$  \\       
1-1 S(13) &  4.0675 & $22\,516.4$   & $1.38 \times 10^{-6}$  & 01 & $ 2.39 \times 10^{-4}$ &  13.7     & 0.33 & $6.17 \times 10^{15}$  \\
          &         &           &                        & 02 & $ 1.80 \times 10^{-4}$ &   8.7     & 0.33 & $4.55 \times 10^{15}$  \\
0-0 S(12) &  3.9968 & $15\,538.5$   & $1.27 \times 10^{-6}$  & 01 & $ 6.50 \times 10^{-4}$ &  41.5     & 0.34 & $1.78 \times 10^{16}$  \\
1-1 S(14) &  3.9414 & $24\,372.4$   & $1.69 \times 10^{-6}$  & 01 & $ 1.03 \times 10^{-4}$ &   6.5     & 0.35 & $2.09 \times 10^{15}$  \\
0-0 S(13) &  3.8464 & $17\,437.7$   & $1.62 \times 10^{-6}$  & 01 & $ 1.51 \times 10^{-3}$ &  78.2     & 0.36 & $3.17 \times 10^{16}$  \\
          &         &           &                        & 02 & $ 1.40 \times 10^{-3}$ &  92.7     & 0.36 & $2.94 \times 10^{16}$  \\
1-1 S(15) &  3.8404 & $26\,257.2$   & $2.00 \times 10^{-6}$  & 01 & $ 1.09 \times 10^{-4}$ &   7.2     & 0.36 & $1.85 \times 10^{15}$  \\
          &         &           &                        & 02 & $ 1.51 \times 10^{-4}$ &  10.8     & 0.36 & $2.57 \times 10^{15}$  \\
1-0 O(7)  &  3.8075 &  8364.9   & $1.06 \times 10^{-7}$  & 01 & $ 1.43 \times 10^{-3}$ &  99.3     & 0.37 & $4.57 \times 10^{17}$  \\
1-1 S(16) &  3.7602 & $28\,199.5$   & $2.32 \times 10^{-6}$  & 01 & $< 1.51 \times 10^{-5}$ &   $< 3$    & 0.38 & $< 2.19 \times 10^{14}$  \\
2-1 O(6)  &  3.7236 & $13\,150.2$   & $2.28 \times 10^{-7}$  & 01 & $< 3.72 \times 10^{-4}$ &$< 15.3 $   & 0.38 & $< 5.47 \times 10^{16}$  \\
0-0 S(14) &  3.7245 & $19\,408.7$   & $2.41 \times 10^{-6}$  & 01 & $< 3.72 \times 10^{-4}$ &$< 15.3 $   & 0.38 & $< 5.18 \times 10^{15}$  \\
1-1 S(17) &  3.6979 & $30\,156.2$   & $2.64 \times 10^{-6}$  & 01 & $ 6.21 \times 10^{-5}$ &   3.8     & 0.39 & $7.87 \times 10^{14}$  \\
3-2 O(5)  &  3.6630 & $17\,811.7$   & $3.52 \times 10^{-7}$  & 01 & $< 3.73 \times 10^{-5}$ & $< 3$    & 0.39 & $< 3.53 \times 10^{15}$  \\
0-0 S(15) &  3.6263 & $21\,408.6$   & $2.41 \times 10^{-6}$  & 01 & $ 6.15 \times 10^{-4}$ &  29.9     & 0.40 & $8.48 \times 10^{15}$  \\ 
          &         &           &                        & 02 & $ 6.49 \times 10^{-4}$ &  31.2     & 0.40 & $8.94 \times 10^{15}$  \\ 
0-0 S(16) &  3.5470 & $23\,451.6$   & $2.83 \times 10^{-6}$  & 01 & $ 2.81 \times 10^{-4}$ &   7.2     & 0.42 & $3.28 \times 10^{15}$  \\ 
1-0 O(6)  &  3.5007 &  7583.7   & $1.50 \times 10^{-7}$  & 01 & $ 8.37 \times 10^{-4}$ &  54.5     & 0.43 & $1.84 \times 10^{17}$  \\ 
0-0 S(17) &  3.4856 & $25\,537.8$   & $3.26 \times 10^{-6}$  & 01 & $ 3.35 \times 10^{-4}$ &  23.1     & 0.44 & $3.39 \times 10^{15}$  \\ 
2-1 O(5)  &  3.4384 & $12\,550.2$   & $3.18 \times 10^{-7}$  & 01 & $< 4.31 \times 10^{-4}$ &$< 22.8$ & 0.46 & $< 4.51 \times 10^{16}$  \\ 
0-0 S(18) & 3.4384 & $27\,638.4$   & $3.68 \times 10^{-6}$  & 01 & $< 4.31 \times 10^{-4}$ &$< 22.8$ & 0.46 & $< 3.89 \times 10^{15}$  \\ 
0-0 S(19) &  3.4039 & $29\,767.7$   & $4.08 \times 10^{-6}$  & 01 & $ 1.62 \times 10^{-4}$ &  15.1     & 0.48 & $1.34 \times 10^{15}$  \\
          &         &           &                        & 02 & $ 1.13 \times 10^{-4}$ &  17.2     & 0.48 & $9.31 \times 10^{14}$  \\
\noalign{\smallskip}        \hline
\noalign{\smallskip}
\end{tabular}
\label{h2_table}
\end{table*} 

\addtocounter{table}{-1}
\begin{table*}[h]
\caption[]{
-Continued
}
\begin{tabular}{lrrlrcrrrr}
\hline\noalign{\smallskip}
\hline\noalign{\smallskip}
\multicolumn{1}{c}{ } &
\multicolumn{1}{c}{$\lambda$} &
\multicolumn{1}{c}{$E_{\rm u}/k~^{\rm a}$} &
\multicolumn{1}{c}{$A~^{\rm b}$} &
\multicolumn{1}{c}{SWS} &
\multicolumn{1}{c}{$I_{\rm obs}~^{\rm c}$} &
\multicolumn{1}{c}{ } &
\multicolumn{1}{c}{$A_\lambda~^{\rm e}$} &
\multicolumn{1}{c}{$N_{\rm u}~^{\rm f}$} \\
\multicolumn{1}{c}{\raisebox{1.ex}[-1.ex]{line}} &
\multicolumn{1}{c}{[$\rm \mu m$]} &
\multicolumn{1}{c}{[K]} &
\multicolumn{1}{c}{[s$^{-1}$]} &
\multicolumn{1}{c}{AOT} &
\multicolumn{1}{c}{[erg~s$^{\rm -1}$cm$^{\rm -2}$sr$^{\rm -1}$]} &
\multicolumn{1}{c}{\raisebox{1.ex}[-1.ex]{S/N$~^{\rm d}$}} &
\multicolumn{1}{c}{[mag]} &
\multicolumn{1}{c}{[cm$^{-2}$]} \\
\noalign{\smallskip}
\hline\noalign{\smallskip}
3-2 O(4)  &  3.3958 & $17\,380.1$   & $4.87 \times 10^{-7}$  & 01 & $ 4.88 \times 10^{-4}$ &   6.5     & 0.49 & $3.38 \times 10^{15}$  \\
0-0 S(20) &  3.3809 & $31\,898.5$   & $4.45 \times 10^{-6}$  & 01 & $< 1.43 \times 10^{-5}$ &  $< 3$     & 0.50 & $< 1.09 \times 10^{14}$  \\
0-0 S(21) &  3.3689 & $34\,040.8$   & $4.78 \times 10^{-6}$  & 01 & $ 2.47 \times 10^{-5}$ &   4.0     & 0.52 & $1.77 \times 10^{14}$  \\
0-0 S(22) &  3.3663 & $36\,149.7$   & $5.06 \times 10^{-6}$  & 01 & $< 1.91 \times 10^{-5}$ &$< 3$    & 0.52 & $< 1.30 \times 10^{14}$  \\
0-0 S(23) &  3.3718 & $38\,299.5$   & $5.27 \times 10^{-6}$  & 01 & $< 1.91 \times 10^{-5}$ & $< 3$   & 0.51 & $< 1.24 \times 10^{14}$  \\
0-0 S(24) &  3.3876 & $40\,419.6$   & $5.42 \times 10^{-6}$  & 01 & $< 1.42 \times 10^{-5}$ &  $< 3$    & 0.50 & $< 8.88 \times 10^{13}$  \\
0-0 S(25) &  3.4108 & $42\,515.1$   & $5.50 \times 10^{-6}$  & 01 & $< 3.33 \times 10^{-5}$ & $< 3$  & 0.48 & $< 2.03 \times 10^{14}$  \\
          &         &           &                        & 02 & $ 2.42 \times 10^{-5}$ &   4.1     & 0.48 & $1.48 \times 10^{14}$  \\
0-0 S(26) &  3.4417 & $44\,573.2$   & $5.51 \times 10^{-6}$  & 01 & $< 3.27 \times 10^{-5}$ & $< 3$ & 0.46 & $< 1.97 \times 10^{14}$  \\
0-0 S(27) &  3.4855 & $46\, 650.3$  & $5.43 \times 10^{-6}$  & 01 & $< 3.27 \times 10^{-5}$ & $< 3$ & 0.44 & $< 1.82 \times 10^{14}$  \\                                                        
0-0 S(28) &  3.5375 & $48\,640.3$   & $5.28 \times 10^{-6}$  & 01 & $< 3.86 \times 10^{-5}$ & $< 3$   & 0.42 & $< 2.41 \times 10^{14}$  \\
0-0 S(29) &  3.5996 & $50\,619.9$   & $5.04 \times 10^{-6}$  & 01 & $< 4.56 \times 10^{-5}$ & $< 3$  & 0.41 & $< 3.00 \times 10^{14}$  \\
1-0 O(5)  &  3.2350 & 6950.6   & $2.09 \times 10^{-7}$  & 01 & $ 3.24 \times 10^{-3}$ &  221.0    & 0.76 & $6.39 \times 10^{17}$  \\
          &         &           &                        & 02 & $ 3.05 \times 10^{-3}$ &  218.0    & 0.76 & $6.02 \times 10^{17}$  \\
          &         &           &                        & 06 & $ 3.05 \times 10^{-3}$ &  144.0    & 0.76 & $6.02 \times 10^{17}$  \\
2-1 O(4)  &  3.1899 & $12\,094.1$   & $4.41 \times 10^{-7}$  & 01 & $ 1.25 \times 10^{-4}$ &   12.4    & 0.88 & $1.28 \times 10^{16}$  \\
3-2 O(3)  &  3.1637 & $17\,092.3$   & $7.04 \times 10^{-7}$  & 01 & $ 4.08 \times 10^{-5}$ &    3.8    & 0.94 & $2.76 \times 10^{15}$  \\
1-0 O(4)  &  3.0039 &  6471.5   & $2.90 \times 10^{-7}$  & 01 & $ 1.28 \times 10^{-3}$ &  119.0    & 1.11 & $2.32 \times 10^{17}$  \\
          &         &           &                        & 02 & $ 1.20 \times 10^{-3}$ &   70.5    & 1.11 & $2.18 \times 10^{17}$  \\
          &         &           &                        & 06 & $ 1.34 \times 10^{-3}$ &   49.6    & 1.11 & $2.43 \times 10^{17}$  \\
2-1 O(3)  &  2.9741 & $11\,789.1$   & $6.40 \times 10^{-7}$  & 01 & $ 3.43 \times 10^{-4}$ &   24.4    & 1.07 & $2.71 \times 10^{16}$  \\
3-2 O(2)  &  2.9620 & $16\,948.5$   & $1.41 \times 10^{-6}$  & 01 & $< 1.98 \times 10^{-5}$ &  $< 3$    & 1.06 & $< 6.96 \times 10^{14}$  \\
2-1 Q(13) &  2.9061 & $23\,926.4$   & $2.22 \times 10^{-7}$  & 01 & $< 4.37 \times 10^{-5}$ & $< 3$    & 0.95 & $< 8.70 \times 10^{15}$  \\
3-2 Q(7)  &  2.8250 & $20\,861.9$   & $3.58 \times 10^{-7}$  & 01 & $< 4.84 \times 10^{-5}$ & $< 3$    & 0.80 & $< 5.07 \times 10^{15}$  \\
1-0 O(3)  &  2.8025 &  6149.2   & $4.23 \times 10^{-7}$  & 01 & $ 6.17 \times 10^{-3}$ &  388.0    & 0.77 & $5.27 \times 10^{17}$  \\
          &         &           &                        & 02 & $ 5.20 \times 10^{-3}$ &  287.0    & 0.77 & $4.44 \times 10^{17}$  \\
2-1 O(2)  &  2.7862 & $11\,635.2$   & $1.29 \times 10^{-6}$  & 01 & $ 1.11 \times 10^{-4}$ &    6.5    & 0.75 & $3.04 \times 10^{15}$  \\
3-2 Q(5)  &  2.7692 & $19\,092.2$   & $3.98 \times 10^{-7}$  & 01 & $ 9.81 \times 10^{-5}$ &    5.9    & 0.74 & $8.52 \times 10^{15}$  \\
3-2 Q(3)  &  2.7312 & $17\,811.7$   & $4.41 \times 10^{-7}$  & 01 & $< 6.55 \times 10^{-5}$ & $< 3$    & 0.71 & $< 4.94 \times 10^{15}$  \\
1-0 Q(13) &  2.7269 & $18\,977.1$   & $1.61 \times 10^{-7}$  & 01 & $ 1.10 \times 10^{-4}$ &    3.3    & 0.71 & $2.27 \times 10^{16}$  \\
3-2 Q(2)  &  2.7186 & $17\,394.5$   & $4.84 \times 10^{-7}$  & 01 & $< 2.63 \times 10^{-4}$ & $<   8.4 $ & 0.71 & $< 1.79 \times 10^{16}$  \\
2-1 Q(9)  &  2.7200 & $18\,099.5$   & $3.03 \times 10^{-7}$  & 01 & $< 2.63 \times 10^{-4}$ & $<   8.4 $ & 0.71 & $< 2.87 \times 10^{16}$  \\
3-2 Q(1)  &  2.7102 & $17\,092.3$   & $6.86 \times 10^{-7}$  & 01 & $< 5.43 \times 10^{-5}$ &  $< 3$    & 0.70 & $< 2.60 \times 10^{15}$  \\
2-1 Q(8)  &  2.6850 & $16\,876.5$   & $3.22 \times 10^{-7}$  & 01 & $< 5.29 \times 10^{-5}$ & $< 3$  & 0.70 & $< 5.32 \times 10^{15}$  \\
2-1 Q(7)  &  2.6538 & $15\,768.7$   & $3.40 \times 10^{-7}$  & 01 & $ 3.32 \times 10^{-4}$ &   10.3    & 0.70 & $3.13 \times 10^{16}$  \\
1-0 Q(11) &  2.6350 & $15\,725.5$   & $1.87 \times 10^{-7}$  & 01 & $ 3.10 \times 10^{-4}$ &    9.8    & 0.70 & $5.29 \times 10^{16}$  \\
          &         &           &                        & 02 & $ 1.86 \times 10^{-4}$ &    5.6    & 0.70 & $3.17 \times 10^{16}$  \\
2-1 Q(5)  &  2.6040 & $13\,889.7$    & $3.74 \times 10^{-7}$  & 01 & $ 6.54 \times 10^{-4}$ &   17.2    & 0.71 & $5.55 \times 10^{16}$  \\
          &         &            &                        & 02 & $ 3.58 \times 10^{-4}$ &   40.5    & 0.71 & $3.04 \times 10^{16}$  \\
1-0 Q(10) &  2.5954 & $14\,220.6$    & $1.99 \times 10^{-7}$  & 01 & $ 1.55 \times 10^{-4}$ &    9.1    & 0.72 & $2.47 \times 10^{16}$  \\
          &         &            &                        & 02 & $ 1.59 \times 10^{-4}$ &    9.3    & 0.72 & $2.53 \times 10^{16}$  \\
2-1 Q(4)  &  2.5850 & $13\,150.2$    & $2.65 \times 10^{-7}$  & 01 & $ 1.60 \times 10^{-4}$ &    7.8    & 0.72 & $1.91 \times 10^{16}$  \\
2-1 Q(3)  &  2.5698 & $12\,550.2$    & $4.12 \times 10^{-7}$  & 01 & $ 4.42 \times 10^{-4}$ &   13.8    & 0.72 & $3.40 \times 10^{16}$  \\
1-0 Q(9)  &  2.5600 & $12\,816.4$    & $2.12 \times 10^{-7}$  & 01 & $< 8.27 \times 10^{-4}$ & $<  4.8 $   & 0.73 & $< 1.24 \times 10^{17}$  \\
          &         &            &                        & 02 & $< 7.85 \times 10^{-4}$ & $< 59.5  $  & 0.73 & $< 1.17 \times 10^{17}$  \\
2-1 Q(2)  &  2.5585 & $12\,094.1$    & $4.50 \times 10^{-7}$  & 01 & $< 8.27  \times 10^{-4}$ & $< 4.8 $    & 0.73 & $<5.82 \times 10^{16}$  \\
          &         &         &                           & 02 & $< 7.85 \times 10^{-4}$ & $< 59.5 $  & 0.73 & $< 5.53 \times 10^{16}$  \\
2-1 Q(1)  &  2.5510 & $11\,789.1$    & $6.37 \times 10^{-7}$  & 01 & $ 4.62 \times 10^{-4}$ &   19.2    & 0.73 & $2.30 \times 10^{16}$  \\
          &         &            &                        & 02 & $ 4.25 \times 10^{-4}$ &   43.4    & 0.73 & $2.11 \times 10^{16}$  \\
4-3 S(1)  & 2.5415  & $22\,761.0$    & $4.49 \times 10^{-7}$  & 02 & $6.98 \times 10^{-5}$  &   4.5     & 0.74 & $4.93 \times 10^{15}$  \\                                           
1-0 Q(8)$^{\rm h}$  &  2.5278 & $11\,521.5$    & $2.23 \times 10^{-7}$  & 01 & $ 3.28 \times 10^{-4}$ &   10.1    & 0.74 & $4.66 \times 10^{16}$  \\
          &         &            &                        & 02 & $ 3.30 \times 10^{-4}$ &   23.3    & 0.74 & $4.69 \times 10^{16}$  \\
1-0 Q(7)  &  2.5001 & $10\,340.3$    & $2.34 \times 10^{-7}$  & 01 & $ 1.76 \times 10^{-3}$ &   95.2    & 0.76 & $2.38 \times 10^{17}$  \\
1-0 Q(6)  &  2.4755 &  9285.7    & $2.45 \times 10^{-7}$  & 01 & $ 8.49 \times 10^{-4}$ &   29.9    & 0.77 & $1.10 \times 10^{17}$  \\
\noalign{\smallskip}        \hline
\noalign{\smallskip}
\end{tabular}
\label{h2_table}
\end{table*} 

\addtocounter{table}{-1}
\begin{table*}[h]
\caption[]{
-Continued
}
\begin{tabular}{lrrlrcrrrr}
\hline\noalign{\smallskip}
\hline\noalign{\smallskip}
\multicolumn{1}{c}{ } &
\multicolumn{1}{c}{$\lambda$} &
\multicolumn{1}{c}{$E_{\rm u}/k~^{\rm a}$} &
\multicolumn{1}{c}{$A~^{\rm b}$} &
\multicolumn{1}{c}{SWS} &
\multicolumn{1}{c}{$I_{\rm obs}~^{\rm c}$} &
\multicolumn{1}{c}{ } &
\multicolumn{1}{c}{$A_\lambda~^{\rm e}$} &
\multicolumn{1}{c}{$N_{\rm u}~^{\rm f}$} \\
\multicolumn{1}{c}{\raisebox{1.ex}[-1.ex]{line}} &
\multicolumn{1}{c}{[$\rm \mu m$]} &
\multicolumn{1}{c}{[K]} &
\multicolumn{1}{c}{[s$^{-1}$]} &
\multicolumn{1}{c}{AOT} &
\multicolumn{1}{c}{[erg~s$^{\rm -1}$cm$^{\rm -2}$sr$^{\rm -1}$]} &
\multicolumn{1}{c}{\raisebox{1.ex}[-1.ex]{S/N$^{\rm d}$}} &
\multicolumn{1}{c}{[mag]} &
\multicolumn{1}{c}{[cm$^{-2}$]} \\
\noalign{\smallskip}
\hline\noalign{\smallskip}
1-0 Q(5)  &  2.4548 &  8364.9    & $2.55 \times 10^{-7}$  & 01 & $ 3.68 \times 10^{-3}$ &  209.0    & 0.78 & $4.60 \times 10^{17}$  \\
1-0 Q(4)  &  2.4375 &  7583.7    & $2.65 \times 10^{-7}$  & 01 & $ 1.61 \times 10^{-3}$ &   76.9    & 0.79 & $1.94 \times 10^{17}$  \\
1-0 Q(3)  &  2.4237 &  6950.6    & $2.78 \times 10^{-7}$  & 01 & $ 5.52 \times 10^{-3}$ &  176.0    & 0.80 & $6.34 \times 10^{17}$  \\
1-0 Q(2)  &  2.4134 &  6471.5    & $3.03 \times 10^{-7}$  & 01 & $ 1.91 \times 10^{-3}$ &   56.3    & 0.80 & $2.01 \times 10^{17}$  \\
1-0 Q(1)  &  2.4066 &  6149.2    & $4.29 \times 10^{-7}$  & 01 & $ 6.31 \times 10^{-3}$ &  195.0    & 0.81 & $4.71 \times 10^{17}$  \\
\noalign{\smallskip}        \hline
\noalign{\smallskip}
\end{tabular}

$^{\rm a}$ The upper level energies were kindly provided by Roueff (1992, private communication). \\
$^{\rm b}$ The Einstein coefficients are taken from  Turner et
al. (\cite{tur77}) and Wolniewicz et al. (\cite{wol98}). \\
$^{\rm c}$ The upper limit for the intensities is calculated from the $3 \sigma$ flux density noise level at
the respective wavelength, times the width of one resolution element. In a few cases of merged lines
it was not possible to derive the individual line intensities. There the measured intensity
of the combined structure -- although the S/N ratio is $> 3$ -- is also indicated as an upper limit for each
of the components. \\
$^{\rm d}$ Calculated from the RMS noise within $\rm \sim 500 \: km\,s^{-1}$. \\
$^{\rm e}$ From the extinction curve shown in Fig.~\ref{extinction}, 
derived from the H$_2$ lines. \\
$^{\rm f}$ Extinction-corrected upper level column density. \\
$^{\rm g}$ The 1-1 S(9) and 0-0 S(9) lines are merged with CO lines. To derive the
H$_2$ line intensities, the CO lines and the background were fit by 
sine functions plus a first order polynomial, 
whereas the H$_2$ lines were fit by a Gaussian. \\
$^{\rm h}$ The 1-0 Q(8) line is merged with H 16-5 line. The respective
intensities were derived by fitting two gaussian functions to the combined structure.
\label{h2_table}
\end{table*} 


\subsubsection{Contribution from the foreground PDR}

The line emission toward Peak 1 must include some contribution from the
photodissociation region bordering the foreground Orion Nebula H~{\sc
  ii} region.  Garden (\cite{gar86}) produced a $\rm H_2$~1-0~S(1) map
which covers OMC-1, the Trapezium, and the Orion Bar PDR.  Following
Burton \& Puxley (\cite{bur90b}), we estimate that the extended
fluorescent $\rm H_2$ flux should amount to about 5\% of the total $\rm
H_2$ emission toward Peak~1 over the SWS aperture.

For an additional estimate of the expected PDR contribution we can
compare the total H$_2$ luminosity toward Peak 1 to that toward the
Orion Bar, a PDR observed nearly edge-on south-east of the Trapezium. We
did observe the Bar with the ISO-SWS (Bertoldi et al., in prep.), and
find that the total H$_2$ emission here amounts to 0.008~ $\rm
erg~s^{-1}cm^{-2}sr^{-1}$, compared to the 0.28~$\rm
erg~s^{-1}cm^{-2}sr^{-1}$ toward Peak 1.  Since the Bar is the brightest
PDR emission peak in the Orion Nebula, we see that the $\rm H_2$
emission from the PDR toward Peak 1 is probably small compared with the
emission arising from the deeply embedded outflow.

\subsubsection{Excitation of molecular hydrogen}

\begin{figure*}[htb] 
\begin{center}
\includegraphics[width=2.\columnwidth]{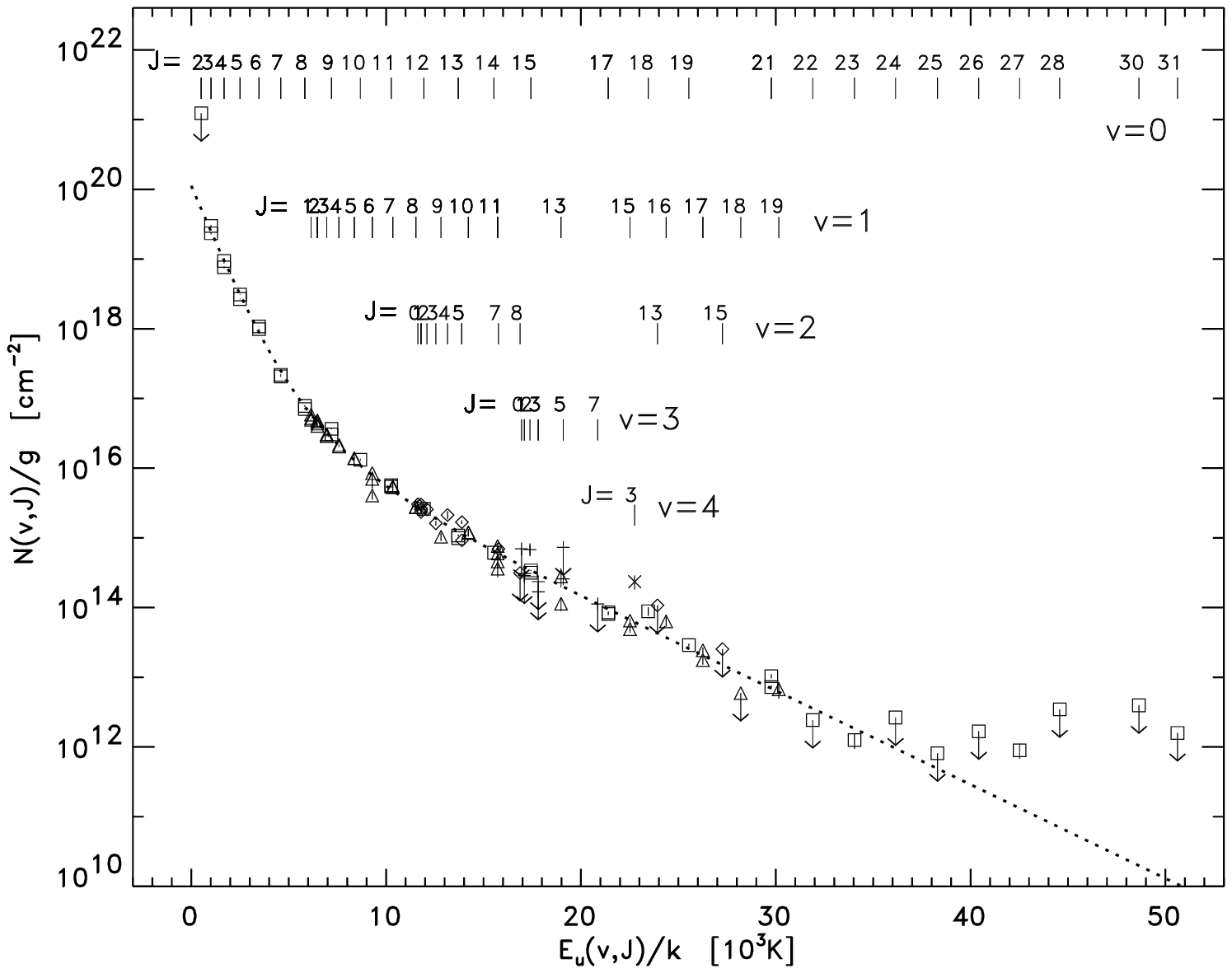}
\caption{Extinction-corrected, observed H$_2$ level column
  densities, divided by their degeneracy, plotted against the upper level
  energy $E(v,J)$. Vibrational levels are distinguished by
  different symbols: squares, triangles, diamonds, $+$, and $\times$
  represent $v=0,$ 1, 2, 3, and 4, respectively. The dotted line represents
  the fit Eq.~\ref{fit}. Error bars represent $1\sigma$ flux
  uncertainties of the line flux integrations,
  but do not include $\sim$30\% calibration uncertainties.
  Some lines were measured in two different AOTs, with the corresponding
  column densities both shown.}
\label{excit}
\end{center}
\end{figure*}

From the line intensities we derived observed column densities of the
levels from which these transitions arise (Eq.\ref{eq:col}).  We correct
these values for extinction with the curve we derived in Sec.~\ref{ex}
(Fig.~\ref{extinction}), to obtain the inherent level column densities
\begin{equation}
N(v,J) = N_{\rm obs}(v,J)~~ 10^{0.4A(\lambda)}.
\end{equation}
The resulting excitation (Boltzmann) diagram is shown in
Fig.~\ref{excit}.

The lack of signs of fluorescent excitation in the level columns
suggest that the molecules might be mostly thermally excited.  An H$_2$
column $N_{\rm H_2,tot}$ in statistical (thermodynamic) equilibrium at a
single kinetic temperature $T$ would yield a level distribution
\begin{equation} 
\label{single_boltz}
 { N(v,J) \over g_J} ~=~  N_{\rm H_2,tot} \;
{ e^{- E(v,J)/k T}\over \sum_{v',J'} g_{J'} \, e^{- E(v',J')/k T} },
\end{equation}
which produces a straight line in the Boltzmann (excitation) diagram
Fig.~\ref{excit}.  An excitation temperature function, $T_{\rm ex}(E)$,
can be assigned to the level distributions at each level energy,
$E(v,J)$, by computing the inverse of the derivative of the line which
best fits $ \ln[N(v,J)/g_J]$ as a function of $E(v,J)/k$.  Near the
lowest energy levels, $T_{\rm ex} \sim 600$~K, whereas at
$E(v,J)/k \geq 14\,000$~K, the excitation temperature rises to $\sim
3200$~K.

To describe the range of excitation temperatures, we decomposed the
distribution of column densities to a sum of five Boltzmann
distributions of different excitation temperatures:
\begin{equation} \label{fit}
  N(v,J)/ g_J = \sum_{i=1}^5  C_i \; e^{- E(v,J)/kT_{{\rm ex},i} },
\end{equation}
where we chose $T_{{\rm ex},i} = (628, 800, 1200, 1800, 3226)$~K, and
the $C_i$ (see Table \ref{column}) were determined by a
least-squares-fit to the observed level columns.  In Fig.~\ref{excit}
the dotted line shows the five-component fit.  From this fit we can also
compute the total warm $\rm H_2$ column density, by summing the column
densities over {\it all} levels following the interpolated level column
distribution:
\begin{eqnarray} \label{n_tot}
  N_{\rm H_2,tot} &  =  & \sum_{v,J} \left[ \frac{N(v,J)}{g_J} \right]
                          \: g_J \nonumber \\ 
                  &  =  & \sum_{v,J} \sum_{i=1}^5 g_J \: C_i \; 
                  e^{-E(v,J)/k T_{{\rm ex},i}} \nonumber \\ 
                  &  =  &  (1.9\pm 0.5) \times 10^{21} \: {\rm cm^{-2}} .
\end{eqnarray}
Adopting a distance of 450~pc (Genzel \& Stutzki \cite{gen89}), this
column corresponds to a warm H$_2$ mass of $(0.06\pm 0.015)~{\rm M}_{\sun}$
within the ISO-SWS aperture. 

By summing from $J = 0$, we extrapolated the observed H$_2$ $(v=0,J\geq 3)$
level populations to the unobserved $(v=0,J = 0, 1, 2)$ levels.
Note that thereby we estimate the total {\it warm} $\rm H_2$ column
density, but we do not account for the {\it total} $\rm H_2$ column
along the line of sight, which includes an additional $\rm \approx
10^{22} \: cm^{-2}$ cold gas from the molecular cloud which embeds the
outflow. Most of this cold $\rm H_2 $ resides in the ground states $J =
0$ and $J = 1$, and does not contribute to the emission observed from the
shock-excited gas in the outflow.

By changing the order of summation in Eq.~\ref{n_tot} we can
compute the column densities corresponding to the
five excitation temperature components,
$N_{{\rm H_2,}i}$ (Table~\ref{column}), such that
\begin{equation} \label{t_components}
   N_{\rm H_2,tot} = \sum _{i=1}^5 \sum_{v,J} g(J) \: 
   C_i \; e^{-E(v,J)/k \, T_{{\rm ex,}i}}
   = \sum _{i=1}^5 N_{{\rm H_2,}i}.
\end{equation}

\begin{table} 
\caption[]{
}
\begin{tabular}{rccr}
\hline\noalign{\smallskip}
\hline\noalign{\smallskip}
\multicolumn{1}{c}{\rule[-2mm]{0mm}{5mm}$T_{{\rm ex,}i}$} & 
\multicolumn{1}{c}{$C_i$} & 
\multicolumn{1}{c}{$N_{\rm H_{\rm 2,i}}$} &
\multicolumn{1}{r}{fraction of}\\
\multicolumn{1}{r}{(K)} &
\multicolumn{1}{c}{(cm$^{\rm -2}$)} &
\multicolumn{1}{c}{(cm$^{\rm -2}$)} &
\multicolumn{1}{c}{warm $\rm H_2$} \\
\noalign{\smallskip}
\hline\noalign{\smallskip}
 628  & $8.80 \times 10^{19}$ & $1.37 \times 10^{21}$ & 72.2\%  \\
 800  & $2.01 \times 10^{19}$ & $3.96 \times 10^{20}$ & 20.9\%  \\
1200  & $3.62 \times 10^{18}$ & $1.07 \times 10^{20}$ &  5.7\%  \\
1800  & $3.83 \times 10^{17}$ & $1.76 \times 10^{19}$ &  0.9\%  \\
3226  & $7.01 \times 10^{16}$ & $6.82 \times 10^{18}$ &  0.4\%  \\
\hline
\noalign{\smallskip}
\end{tabular}
\label{column}
\end{table} 

\begin{figure}[b] 
\begin{center}
\includegraphics[width=1.0\columnwidth]{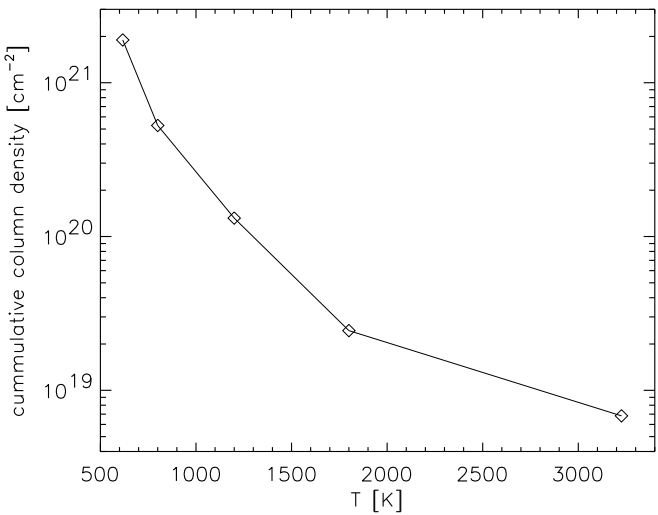}
\caption{
  Cummulative column density of H$_2$ with excitation temperatures larger
  than a given value of $T$, observed within the ISO aperture.
  Derived from the five-component decomposition listed in Table 4.  }
\label{temp}
\end{center}
\end{figure}

Figure~\ref{temp} shows the corresponding
cummulative column density, $N_{\rm H_2}(T_{ex}>T)$, 
plotted against $T$.
With the interpolated excitation distribution Eq.~\ref{fit} the column
densities of all $\rm H_2$ energy levels can be estimated, even those
from which no lines were observed. Then the total H$_2$ rovibrational
emission from the electronic ground state extrapolates to $(0.28 \pm
0.08)$~erg~s$^{-1}$cm$^{-2}$sr$^{-1}$.  Over the ISO-SWS aperture this
amounts to $(17 \pm 5)~\rm L_{\sun}$.  Compared with the total observed
$\rm H_2$ line emission (after extinction correction) of $(0.16 \pm
0.05)$~erg~s$^{-1}$cm$^{-2}$sr$^{-1}$, we find that our line spectra
account for more than half of the total $\rm H_2$ emission.

Our observations target the brightest field in the Orion outflow. The
outflow covers an area of about $2\arcmin \times 2\arcmin$.  The average
H$_2$ brightness over this area we estimate from the 1-0~S(1) map of
Garden (\cite{gar86}) to approximately 20\% of that in our observed
field, so that the total $\rm H_2$ luminosity of the OMC-1 outflow is
estimated to be $(120 \pm 60)~ \rm L_{\sun}$.  This is consistent with the
94~$\rm L_{\sun}$ estimated by Burton \& Puxley (\cite{bur90b}).


\subsubsection{What excites the highest-energy levels?} 
\label{extraexcitation}

Table~\ref{column} and Fig.~\ref{temp} illustrate that only a small
fraction of the warm molecular gas is at the high excitation
temperatures, which reach 3000~K. This is difficult to reconcile with
the expected smooth temperature profile of a single planar C-type shock,
in which the gas temperature changes smoothly, and where a large
fraction of the warm gas is near the maximum temperature (Timmermann
\cite{tim96b}). Even with a distribution of shock speeds, and a
correspondingly wide range in peak temperatures, an excitation
temperature distribution similar to that shown in Fig.~\ref{temp} is
difficult to understand. It would require a velocity distribution where
only a small fraction, about 1\%, of the gas is shocked at the high
speed necessary to produce a 3000 K excitation. In bow shocks, e.g., the
velocity changes slowly with distance from the apex, and such a
distribution of velocities would not be expected.

In dissociative J-type shocks, the molecules are destroyed in the shock,
and they reform in a postshock layer where the temperature has dropped
much below 3000 K, somewhat dependent on the H$_2$ formation rate
efficiency at higher temperatures, which is essentially unknown (e.g.
Bertoldi \cite{ber97}).  Dissociative J-shocks can therefore not
account for the high excitation H$_2$ we observe.

\begin{figure}[htb] 
\begin{center}
\includegraphics[width=1.0\columnwidth]{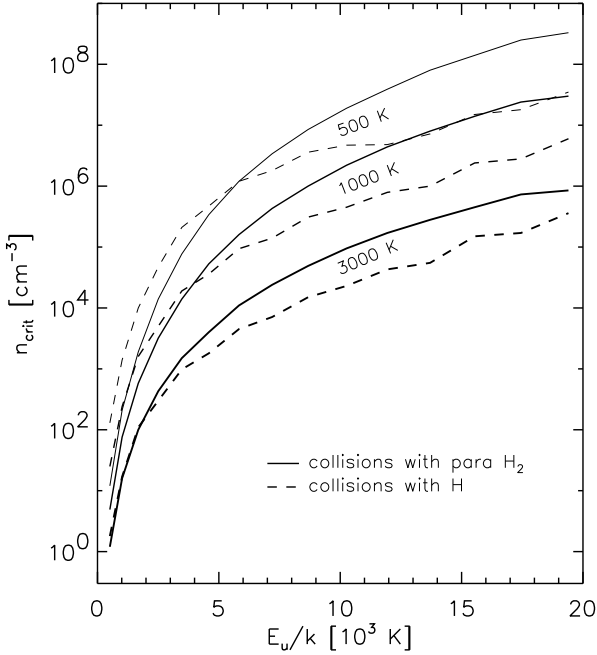}  
\caption{
  Critical densities, $n_{\rm crit}$, of $v=0$ states of H$_2$ as a
  function of level energy, for H$_2$--H$_2$ and H$_2$--H collisions.
  The critical density is the ratio between the sum of all
  radiative rate (Einstein-$A$) coefficients, and the sum of the
  collisional deexcitation rate coefficients from a state to all states
  with lower energy. Einstein-$A$ coefficients were adopted from Turner
  et al.~(1977), and collisional rate coefficients from Bourlot et
  al.~(1999).}
\label{ncrit}
\end{center}
\end{figure}

Even if temperatures of 3000 K or more can be reached in
non-dissociative shocks, the higher H$_2$ levels would remain
subthermally excited unless the gas density is high enough that the
collisional excitation and deexcitation rates are comparable to those
for radiative decay. A ``critical'' gas density can be defined for a
given level as that for which the total collisional deexcitation rate of
this level equals its total radiative decay rate.  In Fig.~\ref{ncrit}
we plot the critical density computed this way for states in the
vibrational ground state, $v=0$, up to $J=16$. We see that even at
kinetic temperatures of 3000 K, gas densities above $10^6\rm cm^{-3}$
would be necessary to maintain the high $v=0$ levels at populations
resembling LTE.  Since such high densities may not prevail in the
shocked gas of the Orion outflow, we may explore mechanisms other than
thermal excitation that could account for the population of the higher
energy states (see also Bertoldi et al.~\cite{ber00}).

\paragraph{Time-dependent C-shocks:} 
When a high velocity outflow strikes dense molecular gas and thereby a
C-type shock is first established, J-type shocks can temporarily form
within the C-shock. In such an embedded J-shock, a small column of gas
is heated to high temperatures (Chi\`{e}ze et al. \cite{chi98}), and if
the density is sufficiently high, this could account for the
high-excitation tail of the column density distribution (Flower \&
Pineau des For\^ets \cite{flo99}).  The lifetime of the embedded J-shock
is small, so that the high-temperature excitation tail would be a
transient phenomenon, unless shocks are constantly reforming. Embedded
J-shocks may also form when a C-shock encounters dense clumps.

\paragraph{Formation pumping:} 
Another possible pumping mechanism of the high-energy states is the
formation of H$_2$. Molecular hydrogen is believed to form on the
surfaces of dust grains.  Some of the 4.5~eV released during the
formation of an H$_2$ molecule is used up to leave the grain, and the
remainder is split between translation, rotation, and vibration of the
new molecule. The exact level distribution of newly formed H$_2$ is
yet unknown, but it could very well contribute to the observed
excitation at intermediate energies, $E\approx 1-3$~eV (Black \& van
Dishoeck \cite{bla87}; Le Bourlot et al. \cite{leb95}).

Using Eq.~\ref{n_tot} we can sum up the column densities of all levels
with energy $E/k \geq 10 \,000$~K, to find a column density $1.30
\times 10^{18}$~cm$^{-2}$, a fraction $6.8 \times 10^{-4}$ of the total
warm H$_2$ column.  Could H$_2$ formation in a steady state produce such
a fraction of molecules in highly excited states?  The pumping rate due
to formation pumping is equal to the H$_2$ formation rate, $n({\rm H})
n_{\rm H} R_{\rm gr}$, where $R_{\rm gr} \approx 5\times
10^{-17}$~cm$^3$~s$^{-1}$ is the H$_2$ formation rate coefficient per
hydrogen nucleus. We estimate the radiative decay rate by starting with
the characteristic radiative lifetime of $\sim 10^6$~sec for a molecule in
a vibrational level $v \approx 5$, and note that $\sim 5$ jumps may be
required to reach the ground state, so that the effective $A$-coefficient
$A_{\rm x} \approx 2 \times 10^{-7}~s^{-1}$.
The population balance for the excited states then writes
\begin{equation}
  R_{\rm gr} ~n_{\rm H}~ n({\rm H}) ~=~ n_{\rm x}({\rm H_2}) ~A_{\rm x}, 
\end{equation}
which yields an excited H$_2$ fraction
\begin{eqnarray} \label{eq:excited}
 {n_{\rm x}({\rm H_2})\over n({\rm H_2})} & = & {n{(\rm H)}\over n({\rm H_2})} {n_{\rm H} R_{\rm gr} \over A_{\rm x} } \nonumber \\
      & = & 5\times 10^{-4} 
      \left(
      {n_{\rm H}\over 10^6{\rm cm^{-3}}}~~
      {n{(\rm H)}\over 2n({\rm H_2})} 
      \right)~,
\end{eqnarray}
which would be consistent with the observed value, if the term in
brackets assumes a value of order unity.  This simple estimate thus
shows that H$_2$ formation could account for some of the high excitation
level populations if the density is high, the atomic fraction not small,
and the formation rate coefficient in the warm shocked gas is somewhat
higher than the value implied at $\approx 100$~K from Copernicus
observations, which is $R_{\rm gr} \approx 3\times 10^{-17}$~cm$^3$~s$^{-1}$
(Jura 1975).

To illustrate the possible importance of H$_2$ formation for the
high-excitation level pumping we show that a simple superposition of two
gas layers with hydrogen nuclei density $n_{\rm H}=10^6\rm cm^{-3}$,
atomic fraction $n({\rm H})/n_{\rm H}=0.5$, and respective temperatures
of 200 K and 800 K, with column densities $N_{\rm H_2}=1.2\times
10^{22}\rm cm^{-2}$ and $1.2\times 10^{21}\rm cm^{-2}$, can in fact
reproduce the observed level column distribution better than any shock
model currently available.  We used the photodissociation front
code of Draine \& Bertoldi
(1996), but without UV illumination, to compute the non-LTE level
distributions for gas at a fixed temperature, density, and molecular
fraction. We include H$_2$ formation with a rate coefficient
$R_{\rm gr}~=~5\times 10^{-17} \rm cm^{3} s^{-1}$ and assume a level
distribution for the newly formed H$_2$ following $N(v,J) \propto (2J+1)
e^{-E(v,J)/kT_{form}}$, with a ``formation temperature''
$T_{form}=5000$~K chosen to match the slope of the observed
high-excitation level distribution.

\begin{figure}[htb] 
\begin{center}
\includegraphics[width=1.05\columnwidth]{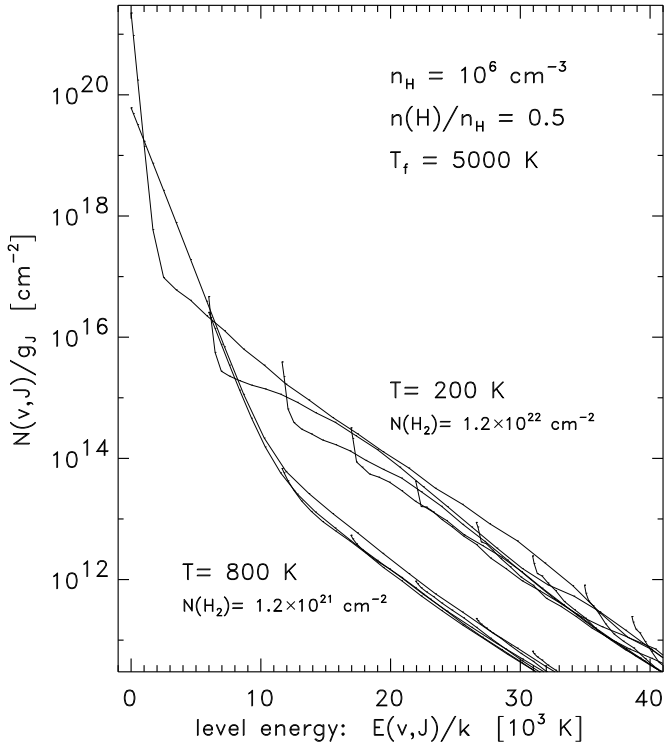}
\caption{
  Level column density distributions for two gas layers at temperatures
  200 K and 800 K with H$_2$ column densities of $1.2\times 10^{22}\rm
  cm^{-3}$ and $1.2\times 10^{21}\rm cm^{-3}$, respectively.  Individual
  vibrational level distributions are shown as separate lines.  The sum
  of both distributions well matches the observed level distribution
  toward Peak 1 which is shown in Fig. 7.}
\label{fig:twoT}
\end{center}
\end{figure}

Fig.~\ref{fig:twoT} illustrates how the newly-formed H$_2$ molecules
give rise to a high-excitation tail in the levels' column density
distribution.  The remaining gas displays a thermal distribution at
least up to the levels which are mainly populated by H$_2$ formation.

\paragraph{Non-thermal collisions:} 
An even more important pumping mechanism for the high-excitation levels
may be non-thermal collisions between molecules and ions in a magnetic
shock.  In magnetic C-type shocks, which are believed to be responsible
for most of the emission in Peak~1, the gas is accelerated through fast
inelastic collisions. In a magnetic precursor the ions, which are tied
to the magnetic field, collide with the undisturbed pre-shock gas at
relative velocities comparable to the shock speed. Such non-thermal
ion--molecule collisions lead to the acceleration of the molecules and to
their internal excitation. High-velocity molecules subsequently collide
with other molecules, leading to a cascade of collisions during which
the relative kinetic energy is in part converted to internal excitation
of the molecules (O'Brien \& Drury \cite{bri96}).  In sufficiently fast
C-shocks, the ion--H$_2$ and H$_2$--H$_2$ collisions can even lead to a
significant collisional dissociation rate.  The molecules dissociated in
a steady-state, partially dissociative shock reform further downstream,
so that across such a shock the H$_2$ dissociation rate equals the H$_2$
reformation rate. For every collisionally dissociated molecule there
will be a larger number of inelastic collisions which did not lead to
dissociation, but to the excitation of the molecule into high
ro-vibrational states, up to the dissociation limit.  The
high-excitation H$_2$ level column densities thereby created should
therefore be larger than those caused by H$_2$ formation alone.

Note that such energetic collision between ions and H$_2$ in C-shocks are
relatively infrequent because ions are rare -- thus the excited
H$_2$ has time enough to cascade to lower levels between collisions,
giving rise to line emission from the highly excited levels. In C-type
shocks, however, dissociations take place too quickly for highly excited
H$_2$ to radiatively decay.

We conclude that non-thermal collisions in partially dissociative
C-shocks could pump the high-excitation states in the H$_2$ electronic
ground state to the levels observed. However, no detailed shock models are
available yet which account for this process.

\paragraph{0 -- 0 S(25):}
The $J=27$ level observed through the $0-0$~S(25) line appears
overpopulated by a factor of seven over what would be expected from the
least-squares fit of the data shown in Fig.~\ref{excit}.  The $J=27$
level is 3.6~eV above ground and only 0.9~eV from the dissociation
limit. H$_2$ molecules which are newly formed on grains are unlikely to
be able to populate states so high, because some fraction of the
formation energy is lost to overcome the grain surface potential, and
some goes to kinetic and vibrational excitation.  Unless we
misidentified the $0-0$~S(25) line, it appears that a different
mechanism may be populating this level and possibly other high levels.
The gas-phase formation of H$_2$ via H$^-$, e.g., might be able to leave
the new molecule in such a high rotational state (Bieniek \&
  Dalgarno \cite{bie79}; Black et al. \cite{bla81}; Launay et al.
  \cite{lau91}).

\subsubsection{Comparison with shock models} \label{model}

For over 20 years, evidence accumulated that the $\rm H_2$ emission from
OMC-1 may arise from shocks (Gautier et al.  \cite{gau76}; Kwan \&
Scoville \cite{kwa76}).  However, the physical nature of these shocks
remains unclear. Models for planar J-type (Hollenbach \& Shull
\cite{hol77}; Kwan \cite{kwa77}; London et al.  \cite{lon77}) or C-type
(Draine \cite{dra80}; Draine \& Roberge \cite{dra82}; Chernoff et al.
\cite{che82}; Draine et al. \cite{dra83}) shocks were unable to
reproduce the observed wide velocity profiles (Nadeau \& Geballe
\cite{nad79}; Brand et al.  \cite{bra89b}; Moorhouse et al.
\cite{moo90}; Chrysostomou et al.  \cite{chr97}), or the wide range of
excitation conditions observed.  Bow shocks were suggested to account
for the observed range of excitation conditions and the wide velocity
profiles (Hartigan et al.  \cite{har87}; e.~g. Smith et al.
\cite{smi91a,smi91b}), but it remains unclear whether these are
predominantly C-type, J-type, or a combination of those.

\begin{figure}[htb] 
\begin{center}
\includegraphics[width=1.\columnwidth]{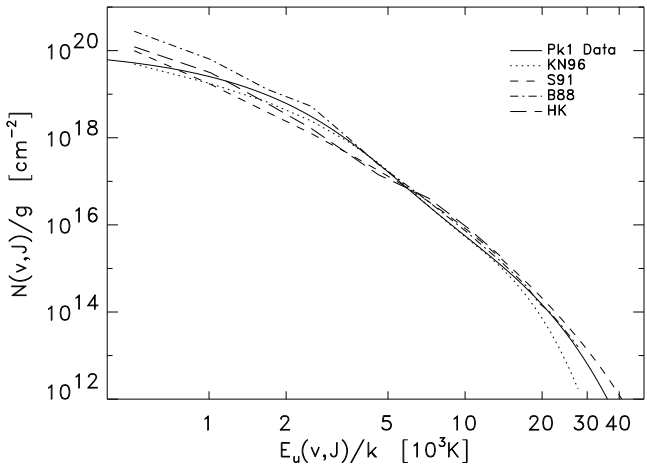}
\caption{
  Comparison of the observed H$_2$ level column distribution (solid
  line) with models: dissociative J-shock plus C-shock model (HK:
  long-dashed), J-type cooling flow model (B88: dash-dotted), bow shock
  model (S91: short-dashed), combination of two planar C-shock models
  (KN96: dotted).}
\label{ex_model}
\end{center}
\end{figure}

\paragraph{Planar shock models:}
In Sect.~\ref{finestructure} we suggested that some fraction of the fine
structure line emission may arise from dissociative J-shocks with
velocities of about 85~km~s$^{-1}$, pre-shock densities $n_{\rm H}\approx
\rm 10^5 - 10^6\,cm^{-3}$, and a beam filling factor of 3--4 (Hollenbach
\& McKee \cite{hol89}).  In such dissociative shocks most of the
excitation of the $v=0$, $ J \leq 5$ levels is collisional, and the
emission arises in the H$_2$ reformation region where the temperature
levels at 400 to 500~K, which is somewhat below the observed excitation
temperature of the lowest levels.  The higher levels would then be
predominantly pumped by newly formed molecules.  Such a model however
neither fits the low excitation nor the higher excitation level
populations very well.  The deficits of the J-shock model could be
compensated if we combined it with a C-shock model, e.g., one of
Kaufman \& Neufeld (\cite{kau96}) with $v_{\rm s}=25$~$\rm km~s^{-1}$,
$n_{\rm H_2}=10^5-10^6~{\rm cm^{-3}}$, and a beam filling factor 0.3.
Such a combined model provides a good fit to the $v=0$, $J=3$ to 9 level
populations, although higher rotational level populations
are predicted too large  (see
Fig.~\ref{ex_model}: HK).

A combination of J-type and C-type shocks would be consistent with the
picture proposed by Chernoff et al.~(\cite{che82}), who suggested that a
high velocity, $\sim {\rm 110~km~s^{-1}}$, wind emanating from an object
near IRc2 drives a $\sim {\rm 30~km~s^{-1}}$ expanding shell of swept-up
material.  The low beam filling factor of the C-shock emission could be
due to the clumping of the ambient medium.

The $\rm H_2$ level populations implied by previous ground-based
observations (e.g. Brand et al. \cite{bra88}; Parmar et al.
\cite{par94}; Burton \& Haas \cite{bur97}) were attempted to match with an
empirical planar J-shock ``cooling flow" model (Brand et al.
\cite{bra88}; Chang \& Martin \cite{cha91}; Burton \& Haas
\cite{bur97}), which assumes that the cooling is dominated by $\rm H_2$,
and cooling by other molecules such as H$_2$O and CO may be neglected.
Such a model can match the medium and high-excitation level populations,
although it somewhat overestimates the population of 
lower rotational levels. These
models assume LTE level distributions, which as we argued above, may not
be a valid assumption for the high-excitation levels if the gas density
is below $10^6\rm cm^{-3}$.  Furthermore, theoretical chemical studies
show that most oxygen not locked in CO is converted to $\rm H_2O$
(Draine et al. \cite{dra83}; Kaufman \& Neufeld \cite{kau96}), which is
detected by ISO-LWS observations toward OMC-1 (Harwit et al.
\cite{har98}; Cernicharo et al. \cite{cer99}). Both H$_2$O and CO should
therefore be significant coolants, and the neglect of this in these
models is worrisome.

Currently available more realistic single shock models do not seem to
fit the observed H$_2$ level distribution.  It appears necessary to
combine at least two shock models, one to account for the
high-excitation level populations, one for the low-excitation levels.
For example, combining two models from Kaufman \& Neufeld
(\cite{kau96}), with shock velocities of 20 and 40 km~s$^{-1}$, and beam
filling factors of 1 and 0.026, respectively, can well match the level
population up to $E/k\approx 20\,000$ K (Fig.~\ref{excit}, KN96); a
pre-shock H$_2$ number density of $\rm 3~\times~10^{5}~cm^{-3}$ was
adopted. The Kaufman \& Neufeld models however do not account for
time-dependency, formation pumping, or non-thermal excitation.

\paragraph{H$_2$ velocity dispersion:}
Optical and near-IR observations with high spectral resolution toward
the OMC-1 outflow show that typical FWHM widths of the H$_2$ lines are
50--60~km s$^{-1}$ (Nadeau \& Geballe \cite{nad79}; Moorhouse et al.
\cite{moo90}; Geballe \& Garden \cite{geb87}; Chrysostomou et al.
\cite{chr97}), and that the line wings can extend to several hundred km
s$^{-1}$ (Ramsey-Howat et al., in prep.). Molecular hydrogen is expected
to be destroyed in shocks with velocities larger than 30 to 50 km
s$^{-1}$, depending on the magnetic field strength.  It is therefore
puzzeling how the H$_2$ emission can show such large velocity
dispersions, even in filaments which are only several arcseconds in
size.

\paragraph{Bow-shocks:}
It has been suggested that the H$_2$ emission arises in bow-shocks,
in which the effective shock velocity decreases from the tip to
the wake. The shock speed at the apex may be high enough to produce a
dissociative J-shock here. But further down the wake, non-dissociative
C-type shocks can prevail, with peak temperatures in the shocked
molecular layers that decrease steadily down the wake.  Thereby a large
range of temperatures for the molecular gas exists in a single bow
shock. This could account for the observed level excitation, and may
also explain the observed constancy of $\rm H_2$ excitation over the
entire OMC-1 outflow (Brand et al.  \cite{bra89a}).

The existence of bow-shocks is also supported by the observation of
double-peaked velocity profiles for isolated regions in the outflow
(Chrysostomou et al. \cite{chr97}), and of knots of [Fe~{\sc ii}]
emission which coincide with ``fingers'' of $\rm H_2$ emission.  Allen
\& Burton (\cite{all93}) suggest that the [Fe~{\sc ii}] and $\rm H_2$
emission trace tips and wakes of bow shocks formed in a stellar outflow.

Recent observations of [Fe~{\sc ii}] and $\rm H_2$~1-0~S(1) velocity
profiles (Tedds et al. \cite{ted99}) however question the
bow shock picture.  Alternatively, Stone et al. (\cite{sto95b}) proposed
that the bows result from Rayleigh-Taylor instabilities when a poorly
collimated outflow accelerates in an ambient medium of decreasing
density, or when catching up with a slower shock.

It is difficult to understand how the high velocity excited H$_2$ can be
produced in a bow shock which is produced, e.g., by a dense bullet which
is moving through a medium initially at rest. Ambient gas which has
passed through parts of the bow shock which are not strong enough to
dissociate the H$_2$ will not be accelerated to velocities much larger
than 30 km s$^{-1}$, unless the magnetic field is very strong. If
alternatively the bow shock arises from a molecular wind impinging on a
dense obstacle, then the problem arises how the molecular wind was
accelerated to over 100 km s$^{-1}$ without destroying the molecules,
and why we do not see a lot more mass, traced by CO, e.g., at such high
velocities.

Smith et al.  (\cite{smi91a}) are able to reproduce the shape and width
of the observed H$_2$ lines in the Orion outflow with bow shock models,
but only by assuming a magnetic field strength of 50 mG, significantly
higher than the 10 mG implied by polarization studies (Chrysostomou et
al. \cite{chr94}).

In Fig.~\ref{ex_model}, we compare our data with a bow shock model by
Smith (\cite{smi91b}), adopting a peak shock velocity of ${\rm
  100~km~s^{-1}}$ and an Alfv\'en speed of ${\rm 2~km~s^{-1}}$.  This
model is able to match the medium-excitation level populations well, but
underestimates the low-excitation, and overestimates the high-excitation
levels.

Note that the H$_2$ excitation in the models of Smith, like for the
planar J-shock model of Brand et al. (\cite{bra88}), was calculated
under the assumption of LTE, and also it ignores non-thermal excitation
mechanisms. These models therefore overestimate the population of the
high energy levels by thermal collision, and at the same time they
underestimate the level population because they neglect non-thermal
excitation mechanisms.

We conclude that current shock models are able to reproduce the overall
H$_2$ level distribution only when combining shocks with a range of
velocities. However, most models do not include the physics most
likely to account for the highest excitation level populations.

\section{Summary}

We obtained spectra with the ISO short wavelength spectrometer of the
2.4--45~$\mu$m emission toward the brightest H$_2$ emission peak of the
Orion OMC-1 outflow. In those spectra we detected a large number of
H$_2$, H~{\sc i}, and atomic/ionic fine structure lines.

\begin{enumerate}
\item Estimating the extinction from relative line intensities we find
  that the atomic hydrogen emission originates in the foreground H~{\sc
    ii} region, whereas the H$_2$ emission comes from the shock-excited
  gas within the star-forming molecular cloud.
\item Most of the atomic fine structure emission originates in the
  foreground H~{\sc ii} region and its bordering photodissociation
  front.
\item The [S~{\sc i}]25\mum\ line and some fraction of the [Si~{\sc
    ii}]34.8\mum\ and [Ne~{\sc ii}]12.8\mum\ emission could arise in
  strong J-shocks.
\item The total warm ($T>$ a few hundred K) H$_2$ column density is
  $(1.9\pm 0.5) \times 10^{21}~\rm cm^{-2}$, and the total warm H$_2$
  mass in the ISO-SWS aperture is $(0.06 \pm 0.015)~\rm M_{\sun}$.  The
  total H$_2$ luminosity within the ISO-SWS aperture is $(17 \pm 5)~\rm
  L_{\sun}$, and when extrapolated to the entire outflow, $(120 \pm
  60)~\rm L_{\sun}$.
\item The H$_2$ excitation reveals no signs of fluorescence
  or a deviation from an ortho-to-para ratio of three.
\item The H$_2$ level column density distribution shows an excitation
  temperature which increases from about 600~K for the lowest rotational
  and vibrational levels to about 3200~K at level energies $E(v,J)/k >
  14\,000$~K.
\item No single steady-state shock model can reproduce the observed
  H$_2$ level populations.  To match both the low- and high-excitation
  level populations, a combination of slow and fast shocks is required,
  or time-dependent magnetic shocks which include transient J-shocks.
  Most shock models lack the processes that are likely to populate
  the high energy H$_2$ levels.
\item The higher energy H$_2$ levels may be excited either thermally in
  non-dissociative J-shocks, through non-thermal collisions between fast
  ions and molecules with H$_2$ in C-shocks, or they are due to newly
  formed H$_2$ molecules.
\item In a most simple model, the overall level distribution of the
  observed H$_2$ is well reproduced by two columns of warm, partially
  dissociated gas at H nuclei density $10^6\rm cm^{-3}$, with respective
  temperatures of 200~K and 800~K and a column density ratio of 10 to 1.
  In this model the high-excitation tail in the H$_2$ level populations
  is due to H$_2$ formation.
\item The highest-excitation line we detected, 0--0 S(25), implies a
  column density of the $v=0,~J=27$ level way above what we would expect
  from the extrapolated excitation of the lower energy levels.
  A different pumping mechanism, such as the gas
  phase formation of H$_2$ via H$^-$, might be responsible for the
  population of the highest energy levels.
\end{enumerate}


\section*{ACKNOWLEDGMENTS}
We are very thankful to C.~Wright, A.~Poglitsch, D.~Lutz, L.~Looney,
M.~Lehnert, and M.~Walmsley for valuable comments, to A.~Schultz for
providing the NICMOS image, to B.~Rubin for providing unpublished
results from his H~{\sc ii} region model, to B.T. Draine for his
contributions to the non-LTE H$_2$ models and his careful reading of the
manuscript, to M.~Smith for providing the H$_2$ line strengths for his
bow shock model, and to the SWS Data Center at MPE, especially to
H.~Feuchtgruber and E.~Wieprecht.

\end{document}